\documentclass[12pt,preprint]{aastex}

\def\e#1{ \times 10^{#1}}
\def\wig#1{\mathrel{\hbox{\hbox to 0pt{%
          \lower.5ex\hbox{$\sim$}\hss}\raise.4ex\hbox{$#1$}}}}
\def\Teff{T_{\rm eff}}
\def\teff{T_{\rm eff}}

\shorttitle{Physical parameters of two very cool T dwarfs}
\shortauthors{Saumon et al.}

\begin{document}

\title{Physical parameters of two very cool T dwarfs}

\author{D. Saumon}
\affil{Los Alamos National Laboratory, MS P365, Los Alamos, NM 87545}
\email{dsaumon@lanl.gov}

\author{M. S. Marley}
\affil{MS 245-3, NASA Ames Research Center, Moffett Field, CA 94035}

\author{S. K. Leggett and T. R. Geballe}
\affil{Gemini Observatory,  670 N. A'ohoku Place, Hilo, HI 96720}

\author{D. Stephens and D. A. Golimowski}
\affil{Department of Physics \& Astronomy, Johns Hopkins University, 
3400 North Charles Street, Baltimore, MD 21218}

\author{M. C. Cushing\altaffilmark{1} and X. Fan}
\affil{Steward Observatory, 933 N. Cherry Avenue, Tucson, AZ 85721}

\author{J. T. Rayner}
\affil{Institute for Astronomy, University of Hawaii,
2680 Woodlawn Drive, Honolulu, HI 96822}

\author{K. Lodders}
\affil{Dept. of Earth \& Planetary Science, Washington University, St. Louis, MO 63130}

\and

\author{R. S. Freedman}
\affil{SETI Institute, MS 245-3, NASA Ames Research Center, Moffett Field, CA 94035}
\altaffiltext{1}{{\it Spitzer Fellow}}

\begin{abstract}

We present new infrared spectra of the T8 brown dwarf 2MASS J04151954--0935066:
2.9--4.1~$\mu$m spectra obtained with the Infrared Camera and Spectrograph
on the Subaru Telescope, and 5.2--14.5$\,\mu$m spectra obtained with the  Infrared
Spectrograph on the {\it Spitzer Space Telescope}. 
We use these data and models to determine an accurate bolometric
luminosity of $\log L_{\rm bol}/L_\odot=-5.67$ and 
to constrain the effective temperature, gravity, mass and age to 725--775$\,$K,
$\log g=5.00$--5.37, $M=33$--58$\,M_{\rm Jupiter}$ and age$=$3--10Gyr.
We perform the same 
analysis using published 0.6--15~$\mu$m spectra for the T7.5 dwarf 2MASS J12171110--0311131,
for which we find a metal-rich composition ([Fe/H]$\sim$0.3), and  $\log L_{\rm bol}/L_\odot=-5.31$,
$\teff=850$--950$\,$K, $\log g=4.80$--5.42, $M=25$--66$\,M_{\rm Jupiter}$ and age$=$1--10Gyr.
These luminosities and effective
temperatures straddle those determined with the same method and models for Gl~570D by \citet{saumon06}
and make 2MASS J04151954--0935066 the coolest and least luminous T dwarf with well-determined properties.
We find that synthetic spectra generated by the models reproduce the observed red through
mid-infrared spectra of 2MASS J04151954--0935066 and 2MASS J12171110--0311131 very well,
except for known discrepancies which are most likely due to the incomplete CH$_4$ opacities.  
Both objects show evidence of departures from strict chemical equilibrium and we discuss
this result in the context of other late T dwarfs where disequilibrium phenomena have been observed. 

\end{abstract}

\keywords{stars: low-mass, brown dwarfs --- 
stars: individual (2MASS J04151954--0935066, 2MASS J12171110--0311131, Gl~570D)}

\section{Introduction}

The Sloan Digital Sky Survey (SDSS; York et al.\ 2000) and the Two
Micron All Sky Survey (2MASS; \citet{bei98,sk06})
have revealed large numbers of ultracool low-mass field dwarfs,
known as L and T dwarfs.  The effective temperatures ($T_{\rm eff}$) of
L dwarfs are $\sim 1450$--2200~K, and those of currently known T dwarfs
are $\sim 700$--1450~K \citep{gol04,vrb04}.  T
dwarfs are identified by the presence of CH$_4$ absorption features in
both the $H$ and $K$ near-infrared bands, the $2\nu_2+\nu_3$ and
$2\nu_3$ features at 1.63 and 1.67$\,\mu$m, the P-branch of the $\nu_2 +
\nu_3$ band at 2.20$\,\mu$m, and the $\nu_3 + \nu_4$ and $\nu_1 + \nu_4$
bands at 2.32 and 2.37$\,\mu$m. Spectral classification schemes using
near-infrared spectra have been presented for T dwarfs by
\citet{bur02,geb02}. These similar schemes, using
primarily the strengths of the H$_2$O and CH$_4$ absorption features at
1--2.5~$\mu$m, were recently merged by \citet{bur06a}.

Currently, there are six dwarfs known that are classified as later than
T7. Five of these were discovered in the 2MASS database: 2MASS~J04151954$-$0935066, 
2MASS~J09393548--2448279, 2MASS~J11145133--2618235,
2MASS~J12171110-0311131 and 2MASS~J14571496$-$2121477 (Burgasser et al.
1999, 2000, 2002, Tinney et al. 2005). The sixth, HD3651B, was recently discovered
through searches for wide companions to stars known to host extrasolar planets
\citep{mu06,luhman06}.
Trigonometric parallaxes have been published for four of these dwarfs
and Table 1 lists these four objects.  Full 2MASS names are given in the Table,
but in the rest of this
paper we use abbreviated names in the form 2MASS~JHHMM$\pm$DDMM, except
for 2MASS~J14571496--2121477 which we refer to by its more commonly used
name of Gl~570D. Table 1 lists published values of trigonometric parallax,
tangential velocity, bolometric luminosity, effective temperature
($T_{\rm eff}$), and surface gravity ($g$).

Three of the very late T dwarfs, 2MASS~J0415--0935, 2MASS~J1217--0311 and
Gl~570D, have been observed spectroscopically in the mid-infrared, and they form the sample for
this paper. Only near-infrared spectra are available for the other three known T7.5--T8 dwarfs
and these will be discussed in a future publication.
Figure 1 plots the red and near-infrared spectra for the three dwarfs in our
sample.  It can be seen that the bands of H$_2$O and CH$_4$ are
extremely strong, as is the K~I 0.77$\,\mu$m doublet. Narrow peaks of
flux emerge from between these bands.

In this paper we present new 2.9--4.1~$\mu$m and 5.5--15~$\mu$m low-resolution 
spectra of one of our sample, 2MASS~J0415--0935 (T8).  The former were obtained
using the Infrared Camera and Spectrograph (IRCS; Kobayashi et al.\ 2000) on
the Subaru Telescope on Mauna Kea, and the latter using the Infrared
Spectrograph (IRS; Houck et al.\ 2004) on the {\it Spitzer Space
Telescope} \citep{wer04}.  We use these data, together with previously
published red and near-infrared spectra and trigonometric parallaxes,
and grids of evolutionary and atmospheric models, to determine
accurate values of $L_{\rm bol}$, $T_{\rm eff}$ and log~$g$ for this dwarf. 
We carry out the same analysis for a second member of the sample, 2MASS~J1217--0311
(T7.5), using published IRS data, and compare these parameters to those
derived in the same way for the third member, Gl~570D (T7.5) by Saumon
et al.  (2006, hereafter S06).  Matches between the data and synthetic spectra
generated from the atmospheric models are
examined.

\section{Observations}

\subsection{Published Observational Data}

Near-infrared spectroscopy and 2MASS photometry for 2MASS~J0415--0935, 2MASS~J1217--0311 
and Gl~570D can be found in the discovery papers (Table 1; Burgasser et al.
1999, 2000, 2002).  Red spectra are presented by Burgasser et al.  (2003).
Near-infrared spectra are also presented by Geballe et al. (2001, 2002) for Gl~570D and
2MASS~J1217--0311, and by Knapp et al. (2004) for 2MASS~J0415--0935.
Near-infrared spectral indices and the implied spectral types can be found for them in
\citet{bur06a}.  Near-infrared photometry on the MKO system \citep{tol02}
is available for Gl~570D (Geballe et al. 2001), 2MASS~J1217--0311
(Leggett et al. 2002), and 2MASS~J0415--0935 (Knapp et al. 2004).

Longer wavelength data are also available for our sample.
Golimowski et al. (2004) present $L^{\prime}_{\rm MKO}$ photometry for all
three dwarfs, and $M^{\prime}_{\rm MKO}$ photometry for 2MASS~J0415--0935.  Patten
et al. (2006) give $Spitzer$'s Infrared Array Camera (IRAC; Fazio et
al.\ 2004) photometry at 3.55, 4.49, 5.73, and 7.87~$\mu$m for the three
dwarfs. IRS 5.2--14.5$\,\mu$m  spectra have been published for Gl 570D
(S06), and 2MASS~J1217--0311 (Cushing et al. 2006). 

\subsection{New 2.9--4.1~$\mu$m IRCS Data}

2MASS~J0415--0935 was observed on the night of 2002 December 12 with the Infrared 
Camera and Spectrograph (IRCS;  Kobayashi et al. 2000) mounted on the 8.2-m Subaru Telescope.  
We used the $L$-band grism with a slit width of 0$\farcs$6 which resulted in a spectrum
covering 2.90 to 4.16~$\mu$m at a resolving power of R$\sim$210.  A series of forty 
35-second exposures were taken in pairs, with the target
at two different positions along the 20$''$ slit.  A nearby A0~V star, HD 29573, 
was also observed to correct for absorption due to the Earth's atmosphere and the 
instrument throughput, and a series of flat-field exposures and dark frames were 
taken at the end of the night.

The spectra were extracted using a modified version of the Spextool data reduction 
package \citep{cu04}.  Each pair of exposures was subtracted and then flat-fielded.  
The spectra were then extracted, and wavelength-calibrated using sky emission
features.  The target spectra were corrected for telluric absorption and the instrument 
throughput by dividing by the spectrum of HD 29573 and multiplying by a Planck function 
for a temperature of 9500$\,$K, as appropriate for an A0~V star.  
The spectrum was rebinned to have 1 pixel per resolution element. The
final signal to noise ratio of the spectrum ranges from 4 to 10 per binned
resolution element.  Flux calibration was obtained by scaling the spectrum to match the 
$L^{\prime}_{\rm MKO}$  magnitude.

\subsection{New 5.2--14.5$\,\mu$m IRS Data}

2MASS~J0415--0935 was observed with the  Infrared
Spectrograph (IRS; Houck et al.\ 2004) on 2005 September 6,
using the {\it Spitzer Space Telescope} as part of our General Observer Cycle 1 campaign to
obtain mid-infrared spectra for a number of late-L and T dwarfs.  The
Short-Low (SL) module was used to obtain both second-order (5.2--7.7$\,\mu$m) 
and first-order (7.4--14.5$\,\mu$m) spectra in 
staring mode.  Two first-order and one second-order sequences were acquired.  
Each sequence
consisted of 18 observations taken at one position along either the
first-order or second-order slit, followed by a nod to a second position
on the slit for an additional 18 observations.  The data were processed
using version 13.2.0 of the IRS pipeline, which removed all of the
``Spitzer-specific" effects and produced a Basic Calibrated Data (BCD)
image for each observation.

The BCD images were reduced using various scripts
and contributed IDL software.  First, a mask identifying bad pixels was
created by flagging those pixels that were variable with time, unstable
or ``Not-a-Number" (NAN) during the course of observations.  The images and mask were then
run through version 1.5 of the IRSCLEAN\_MASK routine (available from the Spitzer
Science Center) to correct the bad pixels.  The sky background in each
image was removed by combining all of the data in the same order, but
other nod position, using a median-rejection algorithm, and subtracting
the resulting sky frame.  The residual background in each subtracted
image was removed by taking the median value of the background on a
row-by-row basis  and then subtracting the value for each row. 
After a final visual inspection of the images, the data were loaded into
the Spectroscopic Modeling Analysis and Reduction Tool (SMART, Higdon et
al.  2004) for extraction and wavelength calibration, using a fixed-width 
aperture of 3 pixels.  All of
the spectra taken in the same order and nod position were combined using
a mean-clipping algorithm.  IRS observations of the standard star HR
6348, taken on September 9 and 14 2005, were similarly combined and
extracted, and a spectral template provided by G. C. Sloan was used to
determine the relative spectral response for each order and nod
position.  These response functions were used to correct for the instrumental
response and to produce an initial flux-calibration for the 2MASS~J0415--0935 
spectra.  The spectra in each order were combined and the two orders merged 
to produce a single spectrum from 5.2 to 14.5$\,\mu$m.  
IRAC Channel 4 photometry from Patten et al. (2006) was used to produce a
final flux calibration --- the spectra were scaled by a constant to produce
a match to the photometry.

The new IRS spectrum of 2MASS~J0415--0935 and the published spectra for Gl
570D and 2MASS~J1217--0311 are shown in Figure 2, with the principal absorbing
species identified. Note that 2MASS~J1217--0311 is located in a part of
the sky having a high mid-infrared background, leading to a somewhat
noisier spectrum than the other two T dwarfs.

\section{Evolution of Brown Dwarfs: Age, Mass, Gravity and Luminosity}

Brown dwarfs cool as they age
with a cooling rate dependent on the mass.  Hence
both mass and age are fundamental parameters for these objects. 
For field brown dwarfs, such as those in Table 1,
constraining the age is a challenge.  In constrast,
HD3651B and Gl~570D are companions to K-type main sequence stars,
which allows age to be constrained based
on kinematics, X-ray luminosity, H$\alpha$ and
Ca~II H and K emission.  \citet{llc06} show
that HD3651B is aged 3--12~Gyr and Gl~570D 1--5~Gyr. Kinematic information only
is available for 2MASS~J0415--0935 and 2MASS~J1217-0311.
These dwarfs have $V_{\rm tan} \sim$ 50--60 km/s, suggesting
that their ages lie in the range 1--10~Gyr, and that
neither is very young, nor are they
likely to be halo members (see for example the discussion of kinematics
in the solar region by Eggen (1998) and references therein).  

A convenient feature of brown dwarfs is that the mass-radius relation
shows a broad maximum at $\sim4\,M_{\rm Jupiter}$, and that their radii 
are within $\pm 30$\% of the radius of Jupiter for masses 0.3--70
M$_{\rm Jupiter}$ and ages $\gtrsim$200~Myr (Figure 3 of \citet{bu01}). 
This peak results from the balance between electron degeneracy pressure,
which dominates in the more massive brown dwarfs, and electrostatic
repulsion of atoms and molecules, which dominates at the lower masses.
With a nearly constant radius over more than a two order of magnitude variation in mass,
the gravity of a brown dwarf is closely correlated to its mass and is approximately 
independent of age for older
brown dwarfs such as those in our sample (Figure 3).  

S06 use the
observed integrated flux and atmospheric and evolutionary models to derive
$\log g= 5.09 - 5.23$ for Gl~570D, corresponding to a mass of
38--47~M$_{\rm Jupiter}$.  \citet{bur06b} use
measured and modelled near-infrared spectral indices to determine $T_{\rm
eff}$ and log~$g$ for a sample of T dwarfs, using the well-determined
values for Gl~570D as calibration of the method; their values are given
in Table 1.  The value for Gl~570D agrees with that determined by S06,
as would be expected. 
Applying the spectral indices technique to HD~3651B, \citet{bur3651} derives
$\log g= 4.7 - 5.1$ and mass 20--40~M$_{Jupiter}$, whereas luminosity arguments
\citep{luhman06,llc06}
imply $\log g= 5.1 - 5.5$ and mass 40--70~M$_{Jupiter}$.
The values of log~$g$ derived by \citet{bur06b} for
2MASS~J1217--0311 and 2MASS~J0415--0935 are consistent with the estimates
made by Knapp et al. (2004) based on $H-K$ color (also shown in Table
1), and suggest masses of 20--30 and 25--35 M$_{\rm Jupiter}$, respectively.  

Here we analyze the brown dwarfs  2MASS~J1217--0311 and 2MASS~J0415--0935
with the same method as described in S06 for Gl 570D, which is based on the
integrated observed flux.
The model atmospheres, spectra and evolutionary sequences that are the 
basis for our analysis of the spectroscopic and photometric data
are described briefly in \S 4.
Section 5 presents the values of $\teff$, $\log g$, luminosity, mass, radius, 
and age we have obtained for 2MASS~J1217--0311 and
2MASS~J0415--0935, along with the values for Gl 570D from S06.
In \S 6, we perform fits of the spectroscopic and photometric data to further constrain the
range of acceptable parameters, including metallicity, and explore the possibility of
non-equilibrium chemistry in their atmospheres.  

\section{Models: Evolution and Spectra}

To constrain the properties of the T dwarfs in our sample with known parallax 
and mid-infrared spectra, we follow the procedure
detailed in S06, \citet{geb01}, and \citet{saumon00}.   The method uses synthetic spectra
to fill in gaps in the spectral coverage and complement the data at long wavelengths,
in order to obtain a bolometric correction.  For objects with measured parallax,  this bolometric correction
is computed self-consistently with evolutionary models which give the radius and the
bolometric luminosity of the brown dwarf. There are no assumptions beyond those that
are implicit in the computation of the atmosphere models, synthetic spectra and evolution.
Because we are studying late T dwarfs, the entire analysis is based on cloudless model atmospheres. 

For solar metallicity, the models used here are the same as those used for Gl 570D by S06.  
The chemical equilibrium calculations for the atmosphere models are described in \citet{lf02}
and use the solar abundances of \citet{lodders03}.
In addition, we have computed a grid of atmosphere models and
associated evolutionary tracks for a non-solar metallicity of $\rm [Fe/H] = +0.3$. 
The non-solar sequences use exactly the same interior physics as the solar sequence; only the 
surface boundary condition, which is extracted from the model atmospheres, is different.

There is growing evidence that the abundances of NH$_3$ and CO in the atmospheres of late T dwarfs
can differ substantially from those expected from chemical equilibrium.
The observed abundances for these two species can be simply explained as departures
from chemical equilibrium driven by vertical mixing in the atmospheres \citep{fl96, lf02, gy99, noll97, saumon00,
saumon03, saumon06}, a phenomenon that has been recognized in giant planets for
some time now \citep{pb77,fl94,bezard02}.  In the lower atmosphere, which is convective, the mixing time
scale over one pressure scale height $H_p$ is very short $\tau_{\rm mix} = H_p/v_{\rm conv} \sim 1\,$s), where
$v_{\rm conv}$ is the convective velocity computed from the  mixing length theory of convection.
Thus, the mixing time scale in the convection zone comes from the mixing length formalism and it
is not adjustable.
In the overlying radiative zone, the physical cause for the mixing is presently unknown and
could arise from the turbulent dissipation of waves generated at the convective/radiative
boundary, meridional circulation, or differential rotation, for example.  Such processes are
expected to mix the radiative atmosphere on much
longer time scales than convection.  For modeling purposes, we treat the
mixing time scale {\it in the radiative zone} as a free parameter, parametrized by the coefficient of
eddy diffusivity, $K_{zz}$.  We consider values in the range $\log K_{zz} {\rm (cm^2\,s^{-1})}= 2$ to 6, which 
corresponds to $\tau_{\rm mix}=H_p^2/K_{zz} \sim 10\,$yr to $\sim 1\,$h, respcetively,  
in the atmospheres of late T dwarfs.
For comparison, $K_{zz}$ would have to be of the order of $10^7$ to 10$^9\,{\rm cm^2\,s^{-1}}$ to get mixing
time scales as short as those found in the convection zone.
Since the IRS spectra and 4--5$\,\mu$m photometry probe a band of NH$_3$ and a band of CO,
respectively, we will consider departures from chemical equilibrium caused by vertical mixing in
comparing our models with the data.

\section{Determination of the Physical Parameters of 2MASS~J0415--0935 and 2MASS~J1217--0311}

Bolometric luminosities have previously been determined for  
Gl~570D, 2MASS~J1217--0311 and 2MASS~J0415--0935 by 
Geballe et al. (2001) and Golimowski et al. (2004).
These authors used the measured parallaxes for these dwarfs and the
observed red through 5~$\mu$m spectral energy distributions,
with an extrapolation to longer wavelengths, to obtain the emitted flux.
The validity of the long-wavelength extrapolation has been verified by
Cushing et al. (2006), who combine {\it Spitzer} mid-infrared
data with shorter wavelength data to determine 
bolometric luminosities for a sample of M1--T5 dwarfs.
Their luminosities agree with those of Golimowski et al. (2004) within
the latter's $\sim$~10\% uncertainties.
Also, S06 repeats the  Geballe et al. (2001) study of Gl 570D, 
adding mid-infrared spectra, and find good agreement with the earlier study.

Here we rederive $L_{\rm bol}$, $\teff$ and $\log g$ for 2MASS~J0415--0935 and 2MASS~J1217--0311
using the method developed in \citet{saumon00} for Gl 229B, which has also
been applied to Gl 570D \citep{geb01,saumon06}.  The method can be applied to
brown dwarfs of known parallax and is not sensitive to systematic 
problems present in the synthetic spectra, as opposed to direct fitting of the 
observed spectra which would be affected by such problems.
For brown dwarfs with a well-sampled spectral energy distribution
($\wig>50$\% of the flux), the method has been shown to be very robust.

We have first summed the observed fluxes
using red spectra \citep{bur03}, near-infrared spectra \citep{geb02,kna04},
3$\,\mu$m spectra where available
(this work) and IRS spectra (this work and Cushing et al. 2006).  
The 1--4$\,\mu$m spectra are flux calibrated using accurate MKO-system $JHKL^{\prime}$ photometry 
from \citet{leg02,gol04,kna04}. 
The IRS spectra are flux calibrated using the IRAC channel 4 photometry from \citet{pat00}.
The uncertainties in the integrated observed flux are dominated by
the flux calibration uncertainties. 
For 2MASS~J0415--0935 they are  5\% at
all wavelengths except for the 1 -- 2.5$\,\mu$m spectrum which has a 3\% uncertainty. 
For 2MASS~J1217--0311 they are 5\% in the red, 3\% in the near-infrared, and 
18\% at IRS wavelengths.  The integrated fluxes at Earth over the observed spectral regions
are $1.532 \pm 0.057 \e{-12}\,$erg$\,$s$^{-1}\,$cm$^{-2}$ for 2MASS~J0415--0935 and
$9.67 \pm 0.55 \e{-13}\,$erg$\,$s$^{-1}\,$cm$^{-2}$ for 2MASS~J1217--0311, which in each case
represents $\sim 75$\% of the bolometric flux.  
The above uncertainties are the sum
of the uncertainties in each integrated segment, and thus are conservative.

Among all possible values of $\teff$ and $\log g$, only a small subset will give a bolometric 
flux correction to the above values that makes $L_{\rm bol}$ consistent with the evolution 
(mass and age) and the known parallax. In the $\{\teff,g\}$ 
space, this subset forms a one-dimensional curve.  The location along the curve can
be parametrized with the mass, age, gravity, effective temperature, luminosity or radius.  
So far, only age constraints and detailed spectral fitting have been used to further
limit the allowed range of the solution along the one-dimensional curve.
For Gl~570D, with an age of  2--5~Gyr and detailed fits of the IRS spectrum, S06 obtained 
$T_{\rm eff}=800-820\,\rm K$, $\log g= 5.09 - 5.23\,\rm cm/s^2$, and 
mass $38 - 47\,\rm M_{\rm Jupiter}$.\footnote{Table 1 of \citet{saumon06} gives erroneous values of
$L_{\rm bol}$ for two of the models.  The correct values are $\log L_{\rm bol}/L_\odot=-5.521$ for model A
and $-5.528$ for model C.}

Applying the same technique of S06, we derive the possible solutions in
$\log g$ and $T_{\rm eff}$ for 2MASS~J0415--0935 and 2MASS~J1217--0311, shown as
curves in Figure 3.
These objects' kinematics constrain their ages to be in the range 1--10~Gyr.
Tables 2 and 3 give a set of specific values along each curve for $T_{\rm eff}$,
$\log g$, luminosity, mass, radius and age for these dwarfs for both [Fe/H]=0
and $+0.3$. Note that the tabulated solutions have nearly constant $L_{\rm bol}$. This is
not an assumption of the method but a consequence of the very weak dependence of 
the calculated bolometric correction on $\teff$ of the atmosphere models when the
bolometric flux correction is $\wig<30$\%.  Our values of $L_{\rm bol}$ agree
with those of \citet{gol04} within the quoted uncertainties for both objects.
For a given object, the solutions for [Fe/H]=0 and $+0.3$ are very close to each other, 
which shows that this method of determination of $(\teff,g)$ is quite robust, given a good sampling of the 
spectral energy distribution (SED).  These models are also
indicated in Figure 3 for solar composition.  The points chosen along each curve correspond to ages
of 1$\,$Gyr (models A \& D), $\sim 3\,$Gyr (models B \& E) and 10$\,$Gyr (models C \& F). 
Atmospheric models and synthetic spectra were generated 
for these parameters, and are compared to the observed spectra and photometry in \S 6 to
determine the optimum range of parameters for each dwarf. 

\section{Comparison of Observed and Synthetic Spectra and Photometry}

With the values of $\teff$ and $g$ constrained to lie along the curves shown in Figure 3
and between the ages of 1 and 10$\,$Gyr, we now seek to further constrain the allowed
range by optimizing the fits of the corresponding synthetic spectra to the observations.
For this purpose, we have three parameters at our disposal.  One is the location along
the solution curve, which  for convenience we will parametrize with the gravity
(mass, radius, age, and $\teff$ are other possible choices), the metallicity of
the atmosphere, for which we consider values of [Fe/H]=0 and $+0.3$, and the time scale
of mixing in the radiative part of the atmosphere, parametrized by the eddy diffusion
coefficient with values of $K_{zz}=0$, 10$^2$, 10$^4$, and 10$^6\,$cm$^2$/s.  Values
of $K_{zz}$ increasing from zero will give increasingly vigorous mixing and larger
departures from chemical equilibrium.  The introduction of vertical mixing and
its associated parameter is motivated by strong evidence of departures from
chemical equilibrium in late T dwarfs which can be well modelled by this process
\citep{noll97,saumon00,gol04,saumon06}; we explore whether this is also
the case for  2MASS~J0415--0935 and 2MASS~J1217--0311.  By combining the
radius, $R(\teff,g)$, from the evolution calculations and the parallax, we compute 
fluxes at Earth from the models and compare them directly with the observed fluxes,
without any arbitrary normalization, except where indicated otherwise. 

In the following comparison of the models with the spectroscopic and photometric data,
we first discuss the near-infrared spectrum, which provides the best constraint on the
metallicity through the $K$-band peak, followed by the mid-infrared spectrum, where 
NH$_3$ bands are indicators of the presence of departures from  chemical equilibrium.
Finally,  3--4$\,\mu$m spectroscopy and photometry provide a consistency
check on the previous determinations and 4--5$\,\mu$m photometry gives a measure of the mixing time scale in 
the radiative region of the atmosphere through the strength of the 4.7$\,\mu$m band of CO.

\subsection{2MASS~J0415--0935}

\subsubsection{0.7--2.5$\,\mu$m spectrum}

The synthetic spectra of each of the triplets of models (Table 2) are very similar to each other in the near-infrared,
because the effect of increasing the gravity is partially canceled by the accompanying increase in effective
temperature from models A to C.  The $(P,T)$ structures of these three models are almost identical,
with deviations of less than 0.14 dex in pressure at a given temperature. 
The general agreement of the synthetic spectra with the data is very good (Figure 4), although
there are some systematic
deviations. For example, (1) the 1.6$\,\mu$m band of CH$_4$ is too weak in the models;
(2) the $J$-band peak is too broad, and (3) the shape of the $Y$-band peak is incorrect.  The first is due
to incompleteness of the CH$_4$ line list at temperatures above 300$\,$K, and the other two to the
absence of CH$_4$ opacity in our model spectra below 1.50$\,\mu$m.  The known bands of CH$_4$ at shorter
wavelengths \citep{strong} have no corresponding line list but they are located such as to
increase the opacity on both sides of the $J$-band peak (Figure 1), making it narrower, and to cut into the blue side
of the $Y$-band peak.  The latter may also be affected by the profile we have adopted
for the far red wing of the K I resonance doublet \citep{bms00}.
As S06 found for Gl 570D, the solar metallicity models
all underestimate the $K$-band flux of 2MASS~J0415--0935, which suggests that it may be enriched
in metals.  Spectra with [Fe/H]=$+0.3$ (models A$^\prime$, B$^\prime$, C$^\prime$ in Table 2) 
reproduce the $K$-band peak very well but the $J$ peak becomes much too strong.
The best synthetic near-infrared spectra  are model C for [Fe/H]=0 and model C$^\prime$ for [Fe/H]=$+0.3$. 
As can be seen from Figure 4, neither value of the metallicity provides a better fit to the
optical and near-infrared spectrum.  No improvement would come from an intermediate value of [Fe/H] either.
The high end of the allowed gravity range is somewhat preferred as it provides a slightly better match to
the $Y$-band peak.  

\subsubsection{5--15$\,\mu$m spectrum}

The mid-infrared spectrum covers strong bands of NH$_3$, a tracer of non-equilibrium chemistry which
can reduce the NH$_3$ abundance in the atmosphere by a factor of $\sim 10$ \citep{saumon06}. 
In this spectral range,
the average flux level decreases noticeably as the gravity increases along the sequence of models 
in Table 2.

We follow the approach of S06 by computing the $\chi^2$ between the synthetic spectra and the observed
spectrum beyond 9$\,\mu$m for each triplet of models, four values of $K_{zz}$ and both metallicities.
In all cases, the best non-equilibrium model ($K_{zz}>0$) provides a much better match than the
best equilibrium model, at a very high level of significance ($\wig>50\sigma$).  If we allow
for arbitrary normalization of the flux rather than considering the absolute fluxes, we find
that the equilibrium abundance of NH$_3$ is still ruled out at the 22$\,\sigma$ level. Therefore
both the average flux level and the detailed shape of the spectrum give a NH$_3$ abundance that is well below
that predicted by equilibrium chemistry.
Other explanations (e.g. low intrinsic nitrogen abundance) are ruled out in S06.
For [Fe/H]=0, the best equilibrium model ($K_{zz}=0$) is clearly A.  The best non-equilibrium model is 
B with $K_{zz}=10^4$ -- $10^6\,$cm$^2\,$s$^{-1}$.  These spectra are shown in Figure 5.
For [Fe/H]=$+0.3$, model A$^\prime$ is 
the best equilibrium model and the best match with non-equilibrium chemistry is obtained with 
model C$^\prime$ and $K_{zz}=10^6\,$cm$^2\,$s$^{-1}$.  Non-equilibrium, solar metallicity models give 
significantly better fits, however.

\subsubsection{2.9--4.1$\,\mu$m spectrum}

The 3--4$\,\mu$m spectrum shows a very strong and very broad 3.3$\,\mu$m CH$_4$ band,
which is saturated at the center (Figure 6).  Only the 3.9 -- 4.1$\,\mu$m region is sensitive to the CH$_4$ abundance,
which depends on the metallicity and potential departures from chemical equilibrium.  The latter act
to reduce the CH$_4$ abundance in the upper atmosphere.
The three solar metallicity models reproduce the spectrum well but underestimate the 3$\,\mu$m bump
by about $2\,\sigma$.  Non-equilibrium chemistry raises the flux in the 3$\,\mu$m bump and 
brings it in excellent agreement with the data (Figure 6).  It also raises the flux longward of 3.9$\,\mu$m
to a level generally higher than the data, but within the $1\,\sigma$ error bars.
The metal-rich model spectra are quite similar but their higher abundance of CH$_4$ 
results in somewhat lower fluxes in the 3.9--4.1$\,\mu$m region than those of the
solar metallicity spectra.  This can be compensated by a higher value of $K_{zz}$, however.
The 2.9--4.1$\,\mu$m spectrum has SNR $\sim 6-7$,
which is not sufficient to discriminate between our different model spectra.  All of them
are in fair agreement with the data; non-equilibrium models (with any $K_{zz}$) fit the 3$\,\mu$m bump 
better, however.

\subsubsection{Photometry}

Figure 7 shows the synthetic photometry in the MKO $L^\prime$ and $M^\prime$  bands and the IRAC bands
[3.55] and [4.49], along with the observed values.  The $M^\prime$ and [4.49] bands both probe the
4.7$\,\mu$m band of CO, a sensitive tracer of non-equilibrium chemistry, which can
dramatically enhance the CO abundance (Figure 6).   

The synthetic $L^\prime$ magnitude is a weak function of gravity, metallicity and $K_{zz}$, with all
models giving values that agree to within $\pm 0.25\,$ magnitude.
With solar metallicity models, the $L^\prime$ magnitude is best reproduced with models A and B with $K_{zz}=0$
and increasing $K_{zz}$ drives the synthetic $L^\prime$ steadily  away from the observed values.
At [Fe/H]=$+0.3$, all three models provide a good match for $K_{zz}\sim 10$ -- 10$^4\,$cm$^2\,$s$^{-1}$.
The MKO $L^\prime$ filter measures the 3.45 -- 4.1$\,\mu$m flux and discriminates better between the
various models than our 2.9 -- 4.1$\,\mu$m spectrum.

Not surprisingly, the IRAC [3.55] magnitude behaves very much like $L^\prime$ but models generally are 
fainter  than the observed value.  The latter can only be reproduced with model C (any $K_{zz}$)
or with any model (including both metallicities) if $K_{zz}\sim10^6\,$cm$^2\,$s$^{-1}$.

The MKO $M^\prime$ and IRAC [4.49] magnitudes behave similarly and give a consistent picture where
$K_{zz}=0$ is strongly excluded and good matches can be obtained for any gravity if $K_{zz}\sim
10^5$ -- 10$^6\,$cm$^2\,$s$^{-1}$ (solar metallicity) or $K_{zz}\sim 10^3$ -- 10$^4\,$cm$^2\,$s$^{-1}$
([Fe/H]=$+0.3$).  This is  strong evidence that the 4.7$\,\mu$m band of CO is deep in 
2MASS~J0415--0935 and that the CO abundance is well above the chemical equilibrium value (Figure 6).

\subsubsection{Optimum models}

No single synthetic spectrum fits this extensive data set but a clear picture emerges.  
The IRS spectrum, and the $M^\prime$ and [4.49] magnitudes all clearly indicate
that NH$_3$ is under abundant and CO is overabundant compared to the values expected
from chemical equilibrium.  In 2MASS~J0415--0935, the chemistry of nitrogen is quenched in the convection zone where the
mixing time scale is determined by the mixing length theory of convection.  The NH$_3$ bands
in the IRS spectrum are therefore good indicators of the presence of non-equilibrium chemistry
but do not provide a handle on the value of the mixing time scale parameter in the 
radiative zone, $K_{zz}$. On the other hand,
the carbon chemistry is quenched in the radiative zone and is quite sensitive to $K_{zz}$.

In this analysis, we are not able to determine the metallicity of 2MASS~J0415--0935.  Both [Fe/H]$=0$ and
[Fe/H]$=+$0.3 
can give equally good agreements with each segment of spectrum and each magnitude (with different
parameters, however).  Overall, most aspects of the SED can be reproduced fairly well by models 
in the range between models B and C (or B$^\prime$ and C$^\prime$ for [Fe/H]=$+0.3$) with a coefficient of eddy 
diffusion of $K_{zz} \wig>10^4\,$cm$^2\,$s$^{-1}$.  The parameters of 2MASS~J0415--0935 are thus:
$\teff=725$--775$\,$K, $\log g=5.00$--5.37, $\log L_{\rm bol}/L_\odot=-5.67$, $M=33$--58$\,M_{\rm Jupiter}$,
and age between 3 and 10$\,$Gyr.  This range of parameters is highlighted in Figure 3.
An example of such a model is shown in Figure 8 and its corresponding synthetic magnitudes are shown in Figure 7.
2MASS~J0415--0935 is the least luminous T dwarf analyzed in details so far and is also the coolest (Figure 3).
In comparison, the $\teff$ and $\log g$ derived by \citet{bur06b} (their Figure 9)
is about 40$\,$K hotter than ours for their quoted range
of gravity ($\log g=4.9$--5.0). Their parameters correspond to $\log L_{\rm bol}/L_\odot=-5.59\pm0.02$,
which is consistent with our model B within their quoted uncertainties.

\subsection{2MASS~J1217--0311}

\subsubsection{0.7--2.5$\,\mu$m spectrum}

For 2MASS~J1217--0311 the near-infrared model spectra for both metallicities again give
very nearly identical solutions for $L_{\rm bol}$, $\teff$ and $\log g$ (Table 3).  And again,
we find that the solar metallicity
models underestimate the $K$-band flux significantly, although the agreement with the
$Y$ and $J$ peaks is excellent (Figure 9).  On the other hand, models with [Fe/H]=$+0.3$ give an excellent fit of
the entire near-infrared SED, except for the usual problem with the 1.6$\,\mu$m band of CH$_4$.  
The maximum $J$-band flux is less sensitive to [Fe/H] at the higher $\teff$ of 2MASS~J1217--0311
than it is for  2MASS~J0415--0935, allowing a good fit of the $K$-band peak at [Fe/H]=$+0.3$
without compromising the good fit of the $J$-band peak.  The width of the $J$-band peak is also
much better reproduced than for  2MASS~J0415--0935 because the temperature where 
it is formed in the atmosphere is higher and the CH$_4$ contribution to the opacity smaller than in  2MASS~J0415--0935.
Water is now mostly responsible for shaping the $J$-band peak and the omission of CH$_4$ opacity
in this wavelength range (\S 6.1.1) is of lesser consequence.  The best
fitting models with [Fe/H]=0 and $+0.3$ are shown in Figure 9.  The quality of the fit with
[Fe/H]=$+0.3$ suggests that 2MASS~J1217--0311 is likely a metal-rich brown dwarf.

\subsubsection{5--15$\,\mu$m spectrum}

The goodness of fit of the models to the IRS spectrum of 2MASS~J1217--0311, as measured by
their $\chi^2$, shows no
statistically significant  sensitivity to the choice of gravity or to the value of
the coefficient of eddy diffusion, $K_{zz}$.  It is not possible to distinguish
between the solar metallicity models D, E and F, or between the metal-rich models
D$^\prime$, E$^\prime$,  and F$^\prime$ (Table 3), or to determine whether the chemistry departs from
strict equilibrium ($K_{zz}=0$).  This is largely due to the lower signal-to-noise
ratio of the spectrum, which ranges from 3 to 12 for $\lambda\ge 9\,\mu$m, compared to 
$\sim 9$ to 54 for the IRS spectrum of 2MASS~J0415--0935.  Since the near-infrared spectrum strongly favors 
a metal-rich composition ([Fe/H]=$+0.3$), we show in Figure 10 the best fitting metal-rich equilibrium model
(model E$^\prime$ with $K_{zz}=0$) and the best fitting non-equilibrium model (model
F$^\prime$) with $K_{zz}$ arbitrarily set to 100$\,$cm$^2\,$s$^{-1}$ since the $\chi^2$
doesn't vary with $K_{zz}$ in this case.   The slope of the spectrum beyond 11$\,\mu$m appears to be
steeper than in the models, which accounts partly for the inability to select among the
various models considered on the basis of the $\chi^2$. The 4.7$\,\mu$m band of CO is 
predicted to be quite strong even for such a modest value of $K_{zz}$.

\subsubsection{Photometry}

In Figure 11, the MKO $L^\prime$ and the IRAC [3.55] and [4.49] magnitudes of 2MASS~J1217--0311 are compared
to the synthetic magnitudes for all six models in Table 3, as the coefficient of
eddy diffusion is varied from $K_{zz}=0$ to 10$^6\,$cm$^2\,$s$^{-1}$.  The metal-rich models
generally fare better in all three bands.  The $L^\prime$ magnitude favors low values
of $K_{zz} \wig< 10^2\,$cm$^2\,$s$^{-1}$, while the IRAC [3.55] band is a rather poor match 
unless $K_{zz}$ is high ($\sim 10^6\,$cm$^2\,$s$^{-1}$).  The IRAC [4.49] band, which measures
the non-equilibrium tracer CO, favors non-equilibrium models with $K_{zz} \wig> 10^2\,$cm$^2\,$s$^{-1}$
and more likely $\wig> 10^4\,$cm$^2\,$s$^{-1}$.
We cannot get a consistent fit for these three magnitudes, however, which most likely reflects 
a limitation of our models. 

\subsubsection{Optimum models}

The near-infrared spectrum of 2MASS~J1217--0311 indicates that it is metal-rich, with
[Fe/H] close to $+0.3$, which is consistent with the photometry shown in Figure 11.  While 
the IRS spectrum shows prominent bands of NH$_3$, it is too noisy
to indicate whether the NH$_3$ abundance departs from chemical equilibrium
or even to constrain the gravity.  The full range of $\teff$ and $\log g$ of
models D$^\prime$, E$^\prime$, and F$^\prime$ (Table 3 and highlighted in Figure 3) 
remains plausible.
Evidence for non-equilibrium chemistry comes solely
from the IRAC [4.49] magnitude, which is 0.2 to 0.4 mag below that expected from
chemical equilibrium.   

To summarize, 2MASS J1217--0311 has [Fe/H]$\sim +0.3$, $\teff= 850$ to 950$\,$K,
$\log g=4.80$ to 5.42, $\log L_{\rm bol}/L_\odot=-5.31$, and we are unable to constrain the
age to better than 1--10$\,$Gyr. This implies a mass between 25 and 65$\,M_{\rm Jupiter}$.
The values derived by \citet{bur06b} agree with our model D$^\prime$ at the $1\sigma$
level, based on their quoted uncertainty.
On the basis of the IRAC [4.49] magnitude, it appears that CO is more abundant than
expected from chemical equilibrium and that a coefficient of eddy diffusion $K_{zz}\wig>10^2\rm cm^2\,s^{-1}$
(possibly as high as 10$^6\,$cm$^2\,$s$^{-1}$) is appropriate to model this object.  A model that represent a good
compromise between all the constraints (model E$^\prime$ in Table 3, with $K_{zz}=10^2\,\rm cm^2\,s^{-1}$)
is compared to the spectrum in Figure 12 and the corresponding photometry is shown in Figure 11.
 
\section{Conclusions}

The new 3--4~$\mu$m and 5--15~$\mu$m spectra presented here for the T8 dwarf
2MASS~J0415--0935, together with its near-infrared spectrum and
trigonometric parallax,  allow a very accurate
determination of its bolometric luminosity ($\log L_{\rm bol}/L_\odot=-5.67\pm0.02$) and 
constrain the possible values of ($\teff,g$).
Although the age for this isolated disk dwarf is not
known, comparisons of synthetic to observed red through mid-infrared spectra show
that we can constrain it to 3--10$\,$Gyr, with corresponding values of 
$T_{\rm eff}=725$--775$\,$K,
$\log g=5.00$--5.37 and a mass between 33 and 58$\,M_{\rm Jupiter}$.  These ranges are
different from error bars in the sense that the values of $\teff$, $\log g$ and $M$ 
within the quoted intervals are correlated (Figure 3 and Table 2).
The metallicity is likely between [Fe/H]=0 and $+0.3$.

Applying the same method of analysis, using published data, 
to the T7.5 dwarf 2MASS~J1217--0311, we find $\log L_{\rm bol}/L_\odot=-5.31\pm0.03$ and an age of 1--10$\,$Gyr.
These correspond to $T_{\rm eff}=850$--950$\,$K,
$\log  g=4.80$--5.42 and a mass of 25 to 65$\,M_{\rm Jupiter}$.   The metallicity of  2MASS~J1217--0311
is [Fe/H]$\sim +0.3$, which indicates that it is very unlikely to be as old as 10$\,$Gyr. The lower end of each of the 
parameter ranges given above is therefore more appropriate.

In terms of $T_{\rm eff}$, 2MASS~J1217--0311 and 2MASS~J0415--0935
straddle the value of 800--820$\,$K derived for the T7.5 dwarf Gl~570D by S06.  
The spectral classifications are uncertain by half a subclass, but on face value
the half subclass difference between Gl~570D and 2MASS~J0415--0935 corresponds
to a $\sim 60\,$K difference in $\teff$ if both objects have the same gravity (Figure 3).
Although 2MASS~J1217--0311
has the same spectral type as Gl 570D, it is warmer by 30--170$\,$K, depending on the respective gravity of
each object. At the low end,
this difference is not significant but a higher value would warrant a study of
the effects of gravity and metallicity on the indices used to determine the spectral type 
of T dwarfs \citep{bur06a}.  

As was found previously for Gl 229B \citep{noll97,oppenheimer98,saumon00} and Gl 570D
\citep{saumon06}, we find evidence for departures from equilibrium chemistry in both objects
studied here.
Specifically, the mid-infrared band of NH$_3$ in 2MASS~J0415--0935 is weaker than expected from
chemical equilibrium models.  The IRS spectrum can be fit very well if the NH$_3$ abundance
is decreased by a factor of 8--10.  The MKO $M^\prime$ magnitude of 2MASS~J0415--0935
is $\sim 1$ magnitude fainter
than equilibrium models predict, a clear indication of a significant enhancement of the CO 
abundance.  Similarly,  the IRAC [4.55] magnitude of  2MASS~J1217--0311 is 0.2 to 0.4 magnitude
fainter than expected, showing a more modest enhancement of CO than in 2MASS~J0415--0935,
compared to the predictions of equilibrium models.

Departures of the chemistry of carbon and nitrogen from chemical equilibrium  is the consequence of
a simple mechanism
based on the kinetics of the relevant chemical reactions and the assumption that vertical mixing takes
place.  Unless the radiative zone is quiescent on time scales of years, non-equilibrium chemistry
of carbon and nitrogen should be a common feature to all low-$\teff$ brown dwarfs. 
These departures from chemical equilibrium can be modelled by considering vertical mixing in the
atmosphere and the time scales of the reactions that are responsible for the chemistry of nitrogen and carbon.  
Deep in the atmosphere, the mixing time scale is determined by the properties of the convection zone,
modelled with the mixing length theory.  In the radiative zone, mixing is caused by some undetermined 
physical process on a longer, and a priori unknown, time scale. This time scale is the only free parameter of 
the non-equilibrium models.   Until the physical processes that can cause mixing in
the radiative zone of  brown dwarf  atmospheres are modelled, we have no indication as to the value of that
mixing time scale. 
\citet{fl94} show that the non-equilibrium abundances of CO, HCN, GeH$_4$, AsH$_3$ and PH$_3$
observed in Jupiter and Saturn can all be modelled very well with $K_{zz}=10^7$ to $10^9\,\rm cm^2\,s^{-1}$.
However, all of these species are quenched deep in the convection zone, which is consistent with
the high values of $K_{zz}$ needed to reproduce their abundances.\footnote{In our models, $K_{zz}$
is used to parametrize the mixing time scale in the radiative zone only and the mixing time scale in
the convection zone is computed with the mixing length theory.  In the planetary science literature,
$K_{zz}$ is used to parametrize the mixing time scale regardless of whether the mixing occurs
in a radiative or a convective zone.}
The value of $K_{zz}$ in
the radiative zones of brown dwarf atmospheres thus remains unconstrained, except for an
upper limit provided by the mixing time scale in the convection zone.  
In this context, the successful modeling of the spectra of brown dwarfs with a simple, one parameter 
model may provide guidance in determining the physical process(es) responsible for the mixing.
As more objects are analyzed in detail, a pattern may emerge in the derived values of $K_{zz}$ 
that can be compared with the results of detailed modeling of mixing
processes when it becomes available. The mixing time scale parametrized by $K_{zz}$ also
enters the modeling of clouds in L dwarfs \citep{ack01}, and empirically determined values of $K_{zz}$
in brown dwarfs allows for the development of a more consistent theory of vertical mixing
and of the cloud model.

Of the four brown dwarfs that have been studied
in enough detail so far, three show clear evidence of non-equilibrium abundances for at least one
of the two key tracers (NH$_3$ and CO): Gl 229B, Gl 570D and 2MASS~J0415--0935.  The fourth,
2MASS~J1217--0311 appears to have a non-equilibrium CO abundance.  In all four objects,
the nitrogen chemistry is quenched in the convection zone, where the mixing time scale is
not adjustable. The depletion of NH$_3$ that
ensues depends very weakly on the value of $K_{zz}$.  On the other hand,
their carbon chemistry is quenched in the radiative zone, and the flux in the 4.7$\,\mu$m band
of CO is very sensitive to the choice of $K_{zz}$.  The values found so far for the coefficient of
eddy diffusivity in the radiative zone are $\log K_{zz}=4-5$
for Gl 229B (based on the CO mole fraction determined by \citet{saumon00}), and $\sim 5-6$ for 2MASS~J0415--0935.
For 2MASS~J1217--0311, the three photometric measurements cannot be fit simultaneously and we find
a weak constraint of $\log K_{zz}>2$. If we (arbitrarily) ignore its $L^\prime$ magnitude, 
then $\log K_{zz}\wig>5$ gives an very good fit to all the other data.  

The present analysis of 2MASS~J0415--0935 and of 2MASS~J12170311 is based on fits of the low-resolution
SED.  Medium resolving power ($\lambda/\Delta\lambda \wig> 2000$)  and especially high resolving power
($\lambda/\Delta\lambda \wig> 20000$) near-infrared spectra open the possibility of
constraining the gravity by fitting  the fine structure of the spectrum, which should
further limit the ranges of $\teff$ and $\log g$ obtained here as well as the
value of $K_{zz}$.  Such an analysis will be greatly simplified by the fact that the
fitted model must lie on the curves shown in Figure 3, which reduces the fit from a 
three-dimensional parameter space ($\teff,\log g,{\rm [Fe/H]}$) to a two-dimensional
space ($\log g,{\rm [Fe/H]}$).

\acknowledgments

This work is based in part on data collected at the Subaru Telescope, which is operated by the     
National Astronomical Observatory of Japan. This work is also 
based in part on observations made with the {\it Spitzer Space
Telescope}, which is operated by the Jet Propulsion Laboratory,
California Institute of Technology under a contract with NASA.
Support for this work was provided by NASA through an award issued by
JPL/Caltech. 
This work was also supported in part under the auspices of the U.S. Department 
of Energy at Los Alamos National Laboratory under Contract W-7405-ENG-36.
MCC acknowledges support from NASA through the Spitzer Space Telescope Fellowship Program,
through a contract issued by JPL/Caltech.  Work by KL is supported by NSF grant AST-0406963.
MSM acknowledges the support of the NASA Office of Space Sciences.
TRG and SKL are supported by the Gemini Observatory,
which is operated by the Association of Universities for Research in
Astronomy Inc. (AURA), under a cooperative agreement with the NSF on behalf of
the Gemini partnership: the National Science Foundation (United States),
the Particle Physics and Astronomy Research Council (United Kingdom),
the National Research Council (Canada), CONICYT (Chile), the Australian
Research Council, CNPq (Brazil) and CONICET (Argentina).

\clearpage

\begin{deluxetable}{clccccccccccc}
\rotate
\tabletypesize{\scriptsize}
\tablecaption{Published Astrometric and Photospheric Data for the T7.5--T8 Dwarfs with Measured Parallax}
\tablewidth{590pt}
\tablehead{
\colhead{Name} &  \colhead{Spectral} & \colhead{Parallax}  & \colhead{$V_{\rm 
tan}$(error)} &   
\colhead{log($L_{\rm bol}/$} & 
\multicolumn{2}{c}{$T_{\rm eff}$~(K)} & &
\multicolumn{2}{c}{log~$g$ (cm/s$^2$)} & &
\multicolumn{2}{c}{References\tablenotemark{f}} \\
\cline{6-7}\cline{9-10}\cline{12-13}
     &\colhead{Type\tablenotemark{a}} & \colhead{(error)(mas)}& \colhead{(km~s$^{-1}$)} & 
 \colhead{$L_{\odot}$)(error)\tablenotemark{b}} & 
\colhead{Spec.\tablenotemark{c}} & \colhead{Lum.\tablenotemark{d}} & &
\colhead{Spec.\tablenotemark{c}} & \colhead{Col./Lum.\tablenotemark{e}} & &
\colhead{Disc.} & \colhead{Astr.} \\
}
\startdata
HD~3651B & T7.5 & 90.03(0.72) & 31.0(1.2) & $-5.60$(0.05) & 760-920 & 780-840 && 4.7-5.1 & 5.1-5.5 && 1,2 & 3 \\
2MASS~J04151954$-$0935066  & T8   & 174.34(2.76) & 61.4(1.0) & $-5.73$(0.05) & 740-760 &  600-750 & &4.9-5.0 & 5.0$\pm$0.25 && 4 & 5 \\
2MASS~J12171110-0311131 & T7.5 & 93.2(2.06)   & 53.5(1.2) & $-5.32$(0.05) & 860-880 & 725-975 &&4.7-4.9 & 4.5$\pm$0.25 && 6 & 5,7 \\
2MASS~J14571496$-$2121477(Gl~570D) & T7.5 & 170.16(1.45) & 56.4(0.9)  & $-5.53(0.05)$ & 780-820 & 800-820 && 5.1 & 5.09-5.23 & & 8 & 3,9 \\
\enddata

\tablenotetext{a}{~Spectral types are based on the near-infrared classification 
scheme for T dwarfs by Burgasser et al.\ (2006). }
\tablenotetext{b}{~Bolometric luminosities from Golimowski et al.\ (2004) and Luhman et al. (2006) }
\tablenotetext{c}{~Derived from the near-infrared spectral ratio technique of 
Burgasser, Burrows \& Kirkpatrick (2006), Burgasser (2006).}
\tablenotetext{d}{~Derived from observed luminosity by \citet{gol04,llc06} and S06.} 
\tablenotetext{e}{~Derived from near-infrared colors by Knapp et al.\ (2004), 
                   or from luminosity by \citet{llc06} and S06.}
\tablenotetext{f}{(1) Mugrauer et al. (2006); (2) Luhman et al. (2006);
(3) ESA (1997)
(4) Burgasser et al. (2002); (5) Vrba et al. (2004);
(6) Burgasser et al. (1999); (7) Tinney et al. (2003);
(8) Burgasser et al. (2000);
(9) van Altena, Lee \& Hoffleit (1995). }
\end{deluxetable}

\begin{deluxetable}{ccccccc}
\tablecaption{Range of Physical Parameters\tablenotemark{a} \ of 2MASS~J0415--0935}
\tablehead{
\colhead{Model} & \colhead{$T_{\rm eff}$~(K)} & \colhead{log~$g$ (cm/s$^2$)} 
& \colhead{log($L_{\rm bol}/L_{\odot}$)} & \colhead{Mass (M$_J$)} 
& \colhead{Radius (R$_{\odot}$)} & \colhead{Age (Gyr)} \\}
\startdata
\multicolumn{7}{l}{[Fe/H]=0} \\
A & 690 & 4.68 & $-5.66$ & 19.5 & 0.104 & 1 \\
B & 730 & 5.04 & $-5.67$ & 34.6 & 0.091 & 3.2\\
C & 775 & 5.37 & $-5.68$ & 57.9 & 0.080 & 10 \\
\multicolumn{7}{l}{[Fe/H]=$+0.3$} \\
A$^\prime$ & 685 & 4.63 & $-5.66$ & 18.3 & 0.106 & 1 \\
B$^\prime$ & 725 & 5.00 & $-5.67$ & 33.4 & 0.093 & 3.2\\
C$^\prime$ & 767 & 5.34 & $-5.68$ & 55.6 & 0.082 & 10\\
\enddata
\tablenotetext{a}{Uncertainties are $\pm 0.02$ dex in $L_{\rm bol}$, which translates to $\pm 9\,$K
              in $\teff$ (at fixed $g$) and $\pm 0.07$ dex on $\log g$ (at fixed $\Teff$,
              or $\pm 0.01$ dex at a fixed age).}
\end{deluxetable}

\begin{deluxetable}{ccccccc}
\tablecaption{Range of Physical Parameters\tablenotemark{a} \ of 2MASS~J1217--0311}
\tablehead{
\colhead{Model} & \colhead{$T_{\rm eff}$~(K)} & \colhead{log~$g$ (cm/s$^2$)} 
& \colhead{log($L_{\rm bol}/L_{\odot}$)} & \colhead{Mass (M$_J$)} 
& \colhead{Radius (R$_{\odot}$)} & \colhead{Age (Gyr)} \\}
\startdata
\multicolumn{7}{l}{[Fe/H]=0} \\
D & 860 & 4.84 & $-5.30$ & 26.4 & 0.100 & 1\\
E & 910 & 5.14 & $-5.31$ & 42.2 & 0.089 & 2.7\\
F & 965 & 5.44 & $-5.31$ & 66.2 & 0.079 & 10\\
\multicolumn{7}{l}{[Fe/H]=$+0.3$} \\
D$^\prime$ & 850 & 4.80 & $-5.31$ & 25.3 & 0.102 & 1\\
E$^\prime$ & 910 & 5.17 & $-5.31$ & 44.7 & 0.089 & 3.2\\
F$^\prime$ & 950 & 5.42 & $-5.32$ & 65.3 & 0.080 & 10\\
\enddata
\tablenotetext{a}{Uncertainties are $\pm 0.03$ dex in $L_{\rm bol}$, which translates to $\sim\pm 25\,$K
              in $\teff$ (at fixed $g$) and $\pm 0.09$ dex on $\log g$ (at fixed $\Teff$,
              or $\pm 0.03$ dex at a fixed age).}
\end{deluxetable}

\clearpage

\begin{figure} \includegraphics[angle=0,scale=.60]{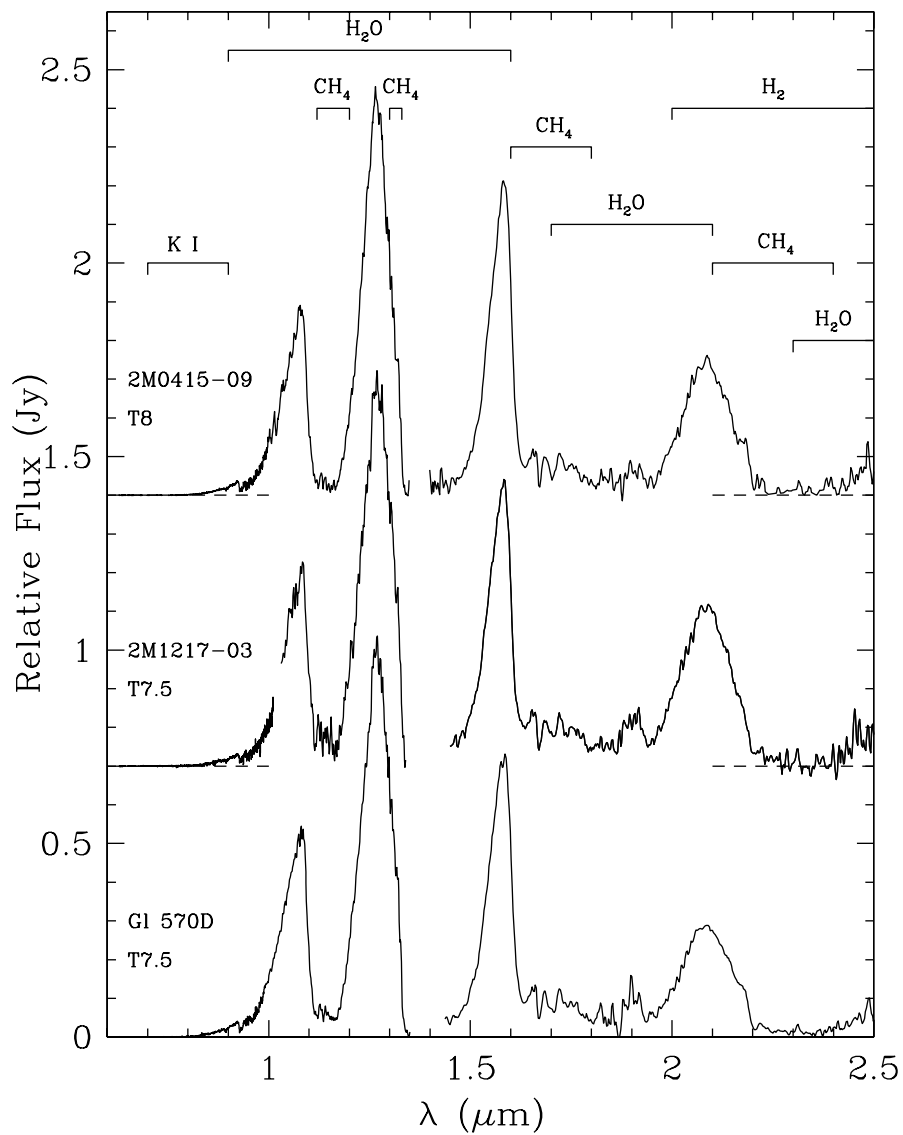} 
\caption{Ground-based red and near-infrared spectra of three of the six known T 
dwarfs later than T7:
2MASS~J0415--0935, 2MASS~J1217--0311 and Gl~570D.
Broad and strong absorption features are identified. The spectra are normalized to 
the peak flux and offset, dashed lines indicate the zero flux level. Data 
sources are Burgasser et al. (2003), Geballe et al. (2001, 2002), and Knapp et al. 
(2004).}
\end{figure}

\begin{figure} \includegraphics[angle=0,scale=.60]{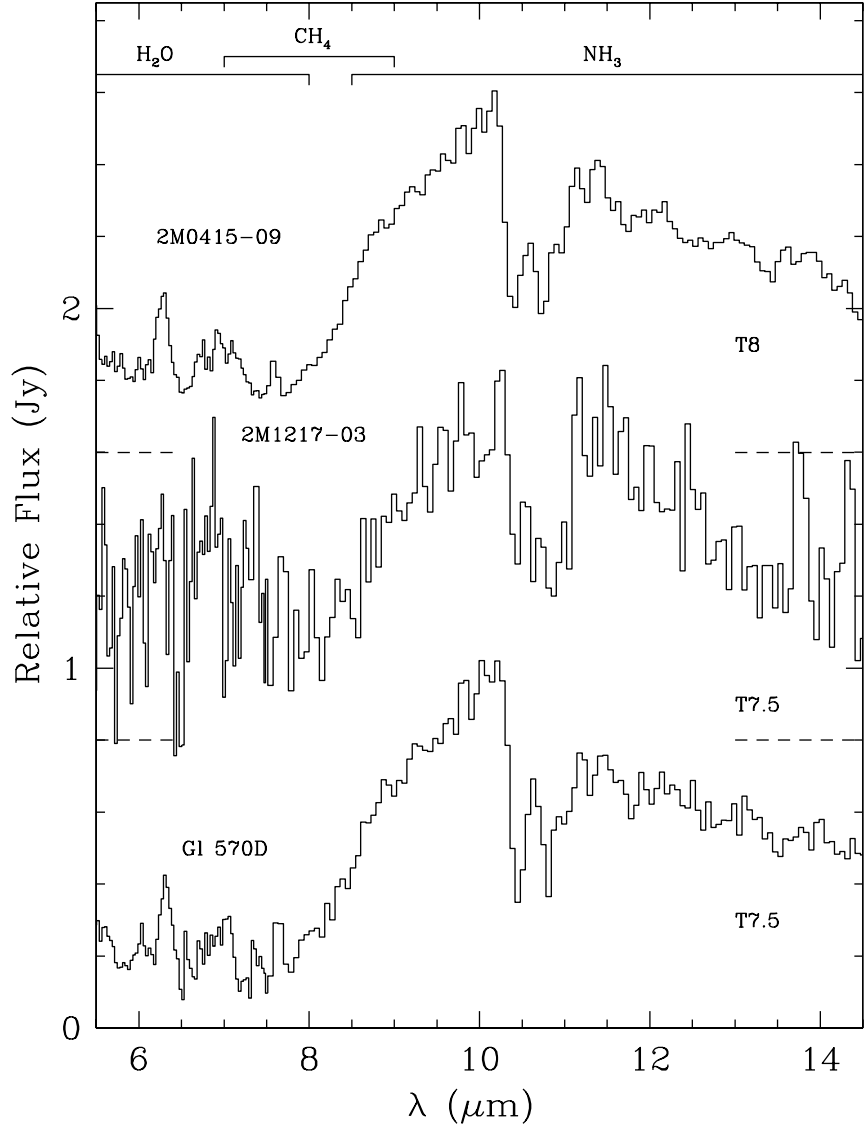} 
\caption{$Spitzer$ IRS spectra of 2MASS~J0415--0935, 2MASS~J1217--0311 and Gl~570D.
Broad and strong absorption features are identified. The spectra are normalized to 
the peak flux and offset, dashed lines indicate the zero flux level.
Data sources are Cushing et al. (2006), S06 and this work.}
\end{figure}

\begin{figure} \includegraphics[angle=0,scale=.80]{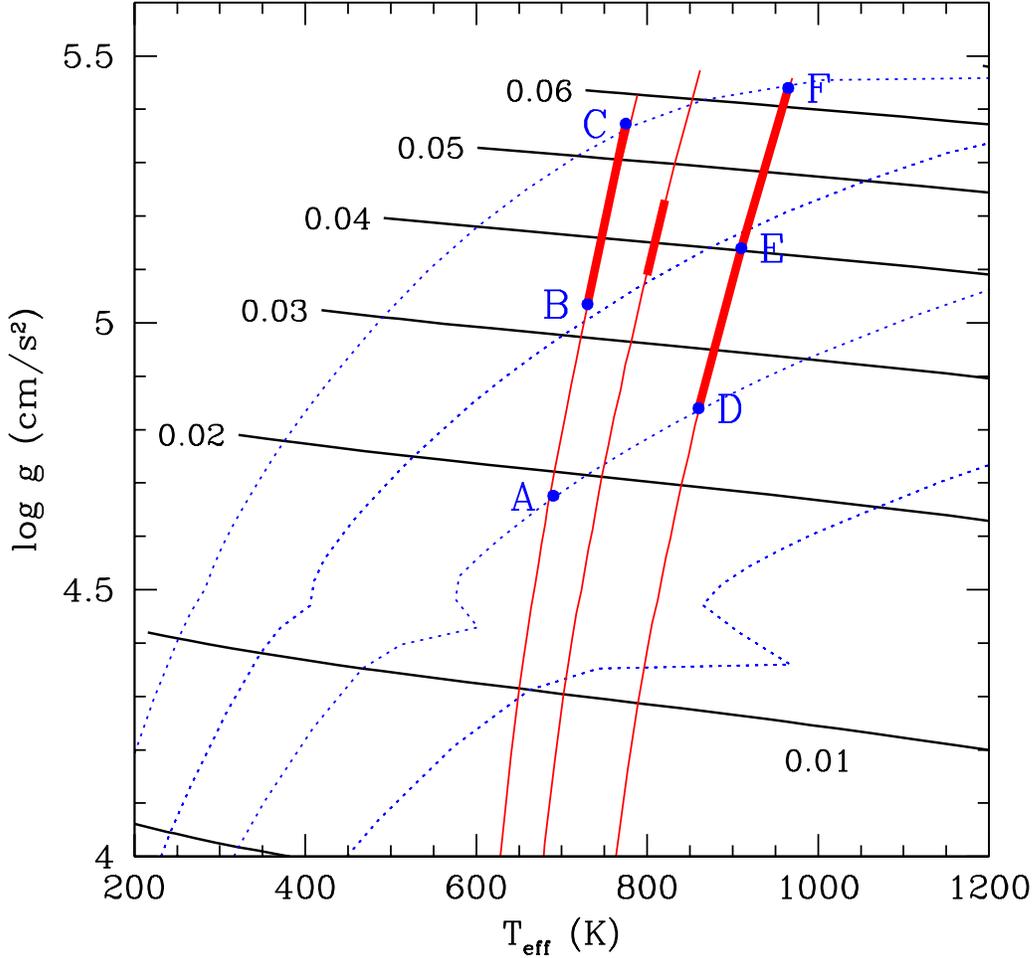}
\caption{Plot of effective temperature vs. surface gravity showing brown
dwarf evolution with [Fe/H]=0.
The allowed ($T_{\rm eff}$, log~$g$) values 
for 2MASS~J0415--0935, Gl~570D, and 2MASS~J1217--0311 fall along the nearly
vertical red lines from left to right, respectively.
$L_{\rm bol}$ is very nearly constant along each curve.
The heavy portion of each curve shows the restricted
range of the most likely solutions based on considerations of age and fitting the 
spectrum and photometry of each object.
Our brown dwarf evolutionary tracks are shown by thick black lines labeled with the 
mass in M$_{\odot}$. Blue dotted lines are isochrones for 0.3, 1, 3 and 10$\,$Gyr (from
right to left) . Filled
blue circles show the three models for 2MASS~J0415--0935 and the three models for
2MASS~J1217--0311 specified in Tables 2 and 3. [{\it See the electronic edition of the 
Journal for a color version of this figure.}]}
\end{figure}

\begin{figure} \includegraphics[angle=0,scale=.80]{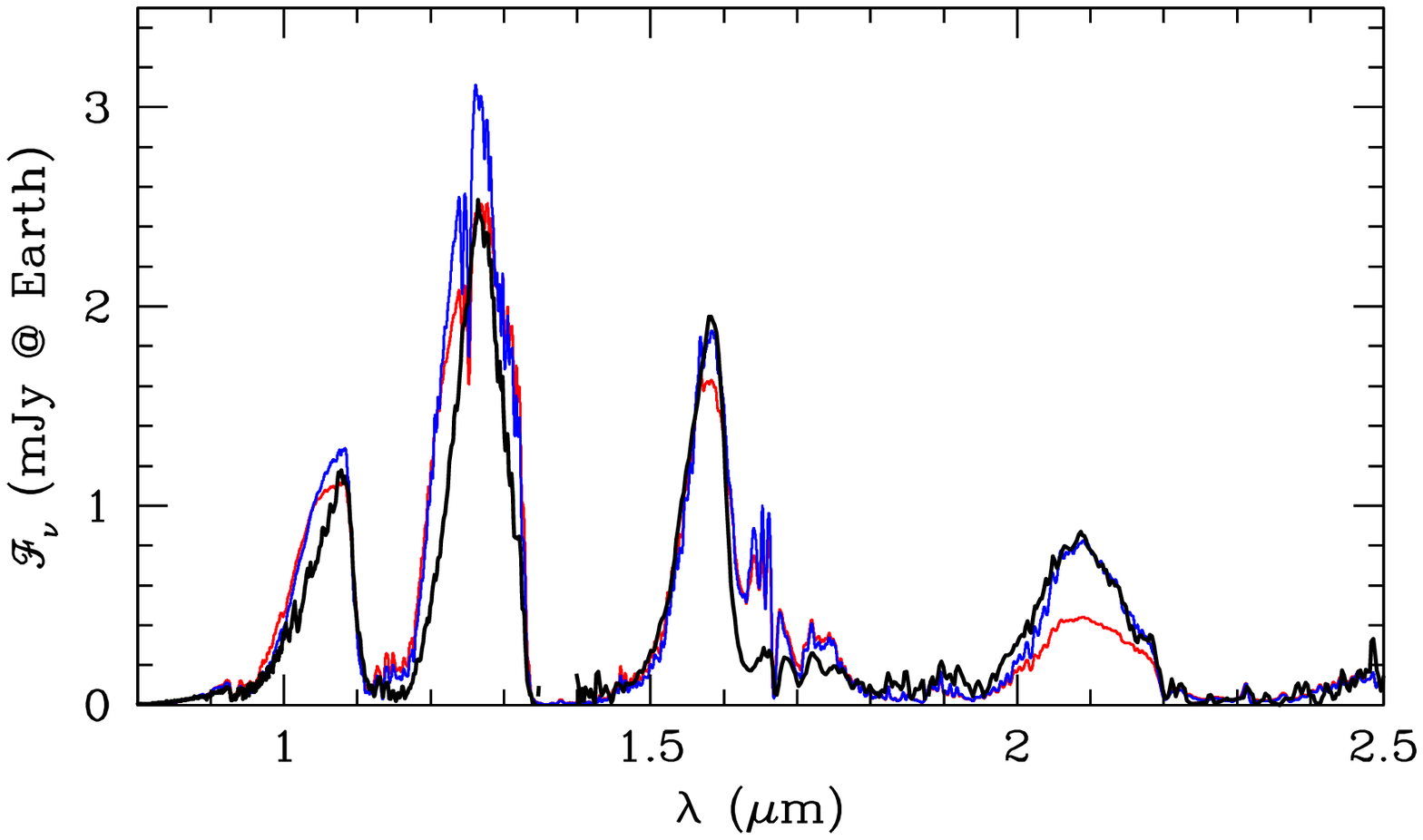}
\caption{Comparison of the optical through near-infrared spectrum of 2MASS~J0415--0935 with models
of different metallicities.  The red curve (lower thin line) is the
best fitting model with [Fe/H]=0 (model C) and the blue curve (upper thin line) is the best fitting
model with [Fe/H]=$+0.3$ (model C$^\prime$). Both models assume chemical equilibrium ($K_{zz}=0$).
The model fluxes have not been normalized to the data but have been smoothed to a resolving power of
$\lambda/\Delta\lambda=500$ to approximate that of the data.  The data is shown by the heavy black line.
[{\it See the electronic edition of the Journal for a color version of this figure.}]}
\end{figure}

\begin{figure} \includegraphics[angle=0,scale=.80]{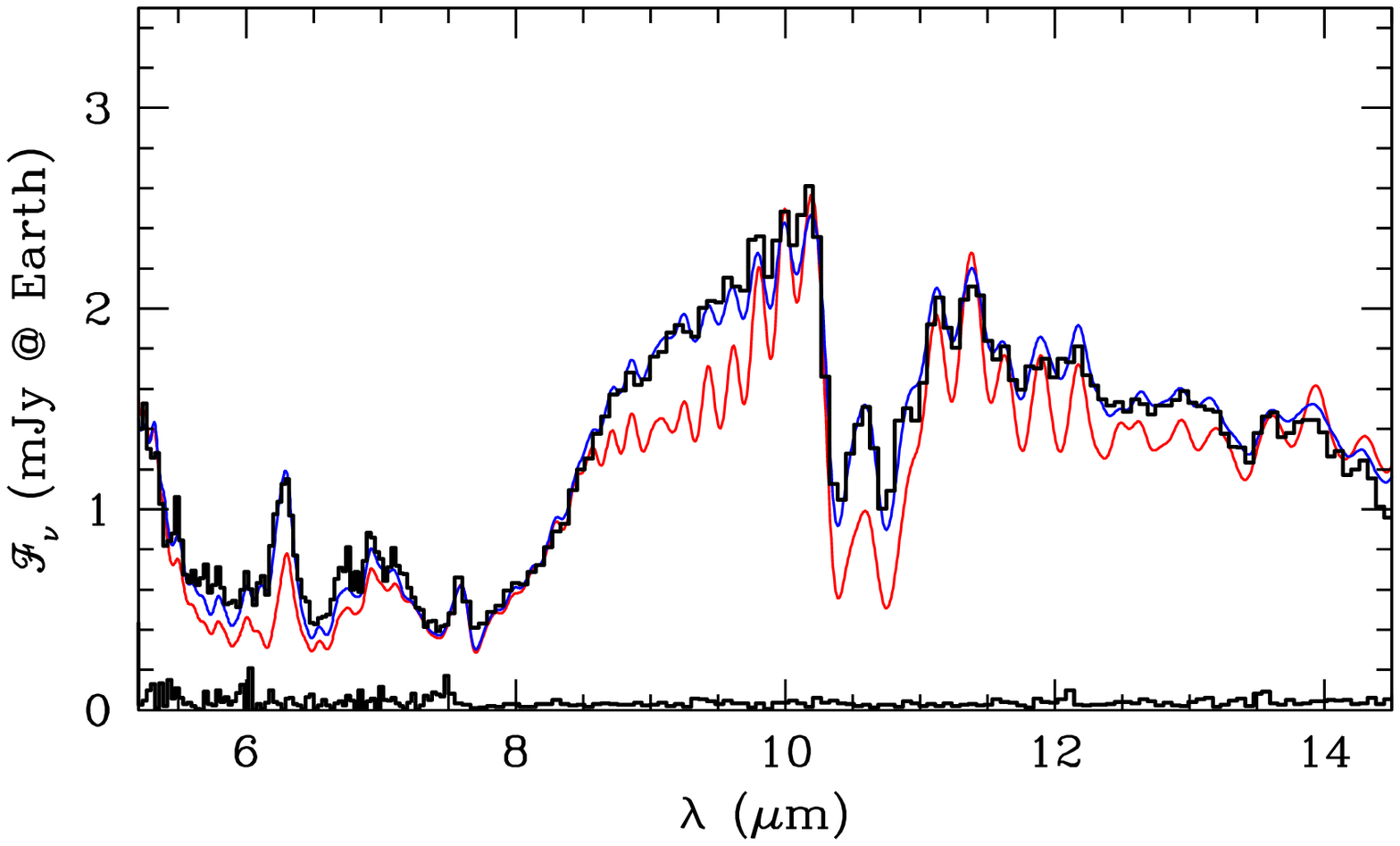}
\caption{Fits of the IRS spectrum of 2MASS~J0415--0935 showing the difference between
a model in chemical equilibrium and a model that includes vertical transport that drives
the nitrogen and carbon chemistry out of equilibrium.  The red curve (lower thin line) is the
best fitting model in chemical equilibrium (model A from Table 2, with 
$K_{zz}=0$) and the blue curve (upper thin line) is the best fitting non-equilibrium model
(model B with $K_{zz}=10^6\,$cm$^2\,$s$^{-1}$).  The data and the noise spectrum
are shown by the histograms (black).  The uncertainty on the flux calibration of the
IRS spectrum is $\pm5$\%.
The model fluxes, which have not been normalized to the data, are shown at the
resolving power of the IRS spectrum ($\lambda/\Delta\lambda=90$).
[{\it See the electronic edition of the Journal for a color version of this figure.}]}
\end{figure}

\begin{figure} \includegraphics[angle=0,scale=.80]{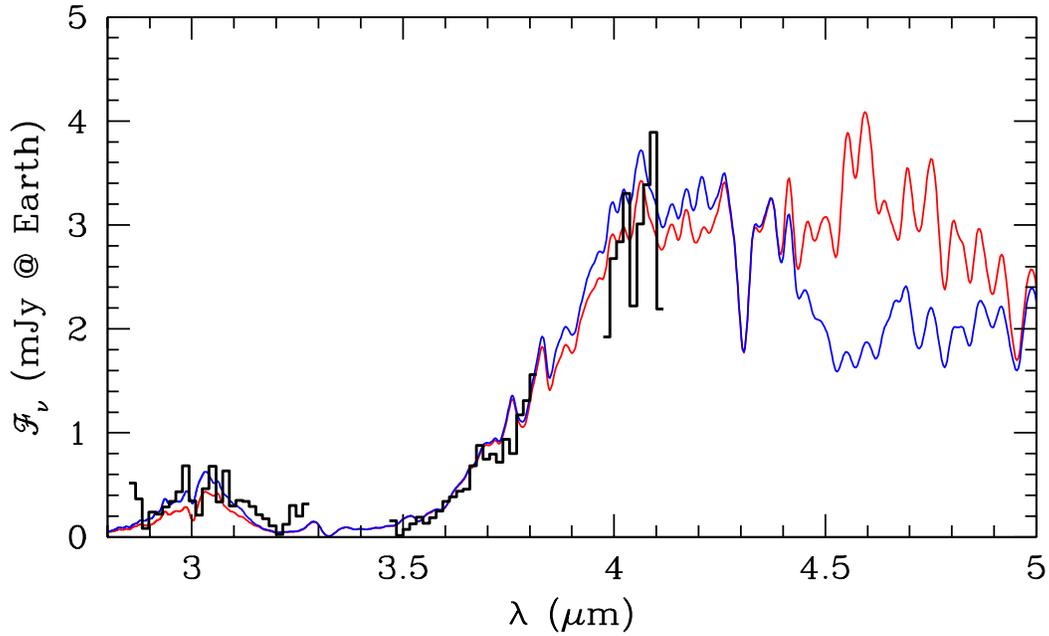}
\caption{Comparison of the 2.9--4.1$\,\mu$m spectrum of 2MASS~J0415--0935 with representative 
models in and out of chemical equilibrium.  The red (dotted) curve is model B with $K_{zz}=0$ (chemical equilibrium) and
the blue curve is model  B with $K_{zz}=10^4\,$cm$^2\,$s$^{-1}$.  The appearance of the
4.7$\,\mu$m band of CO due to non-equilibrium chemistry can bee seen in the models between
4.4 and 4.9$\,\mu$m.  The data is shown by the black histogram.  The model fluxes have not been normalized to the data
and are shown at a resolving power of $\lambda/\Delta\lambda=200$.
[{\it See the electronic edition of the Journal for a color version of this figure.}]}
\end{figure}

\begin{figure} \includegraphics[angle=0,scale=.80]{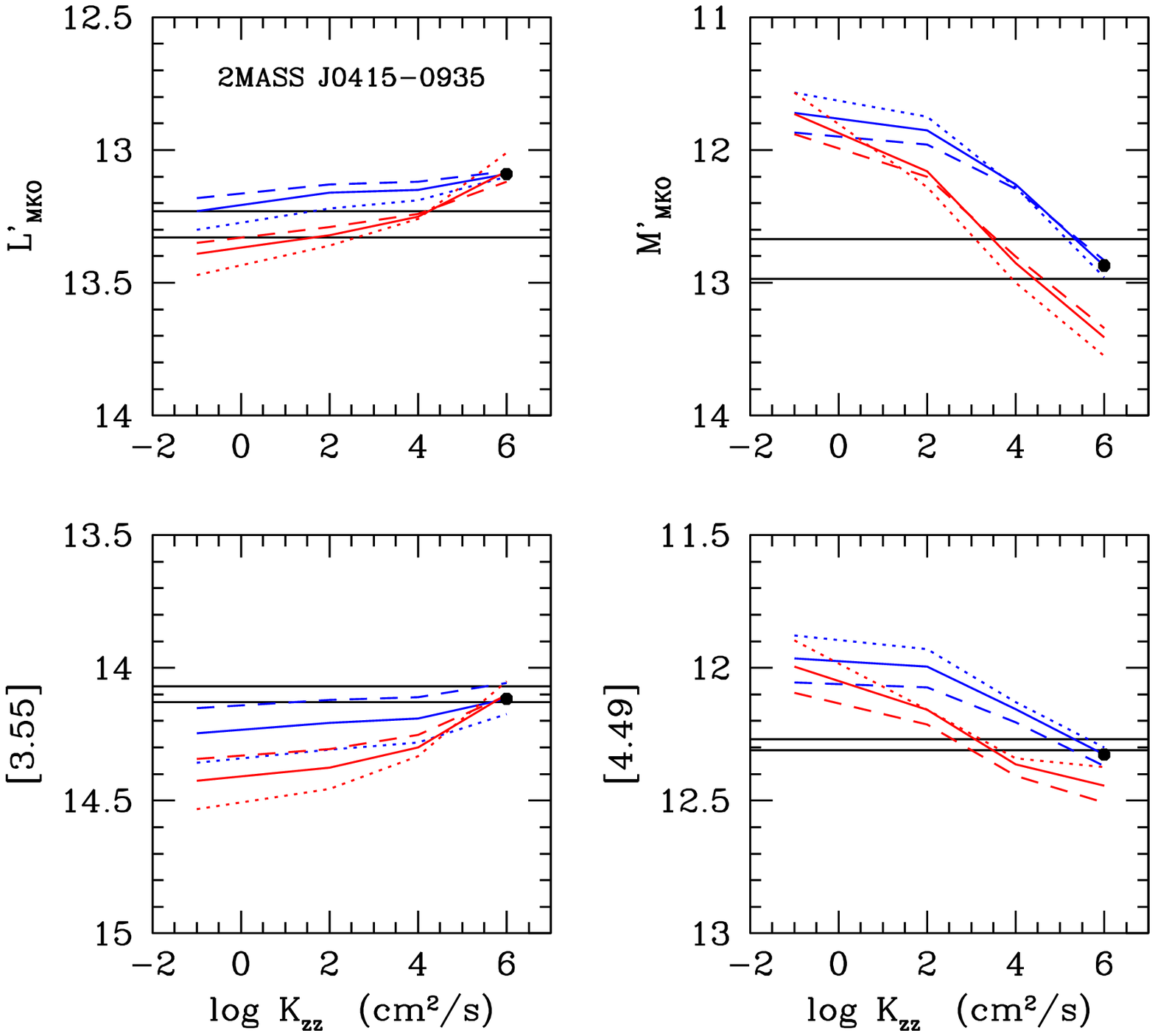}
\caption{MKO and IRAC photometry of 2MASS~J0415--0935 as a function of the coefficient of eddy diffusion
for a range of models.  Observed values ($\pm1\sigma$)
are shown by the horizontal black lines.  Synthetic photometry for each of the models in Table
2 is shown by the various curves: dotted lines: models A and A$^\prime$; solid lines: models
B and B$^\prime$; dashed lines: models C and C$^\prime$.  Blue curves are for [Fe/H]=0
and red curves for [Fe/H]=$+0.3$, the latter are always below the former for a given pair of models.  
The solid dots
show the values for a representative best fit model (see Figure 8).  Models with $K_{zz}=0$
(in chemical equilibrium) are plotted at $\log K_{zz}=-1$.
[{\it See the electronic edition of the Journal for a color version of this figure.}]}
\end{figure}

\begin{figure} \includegraphics[angle=0,scale=.80]{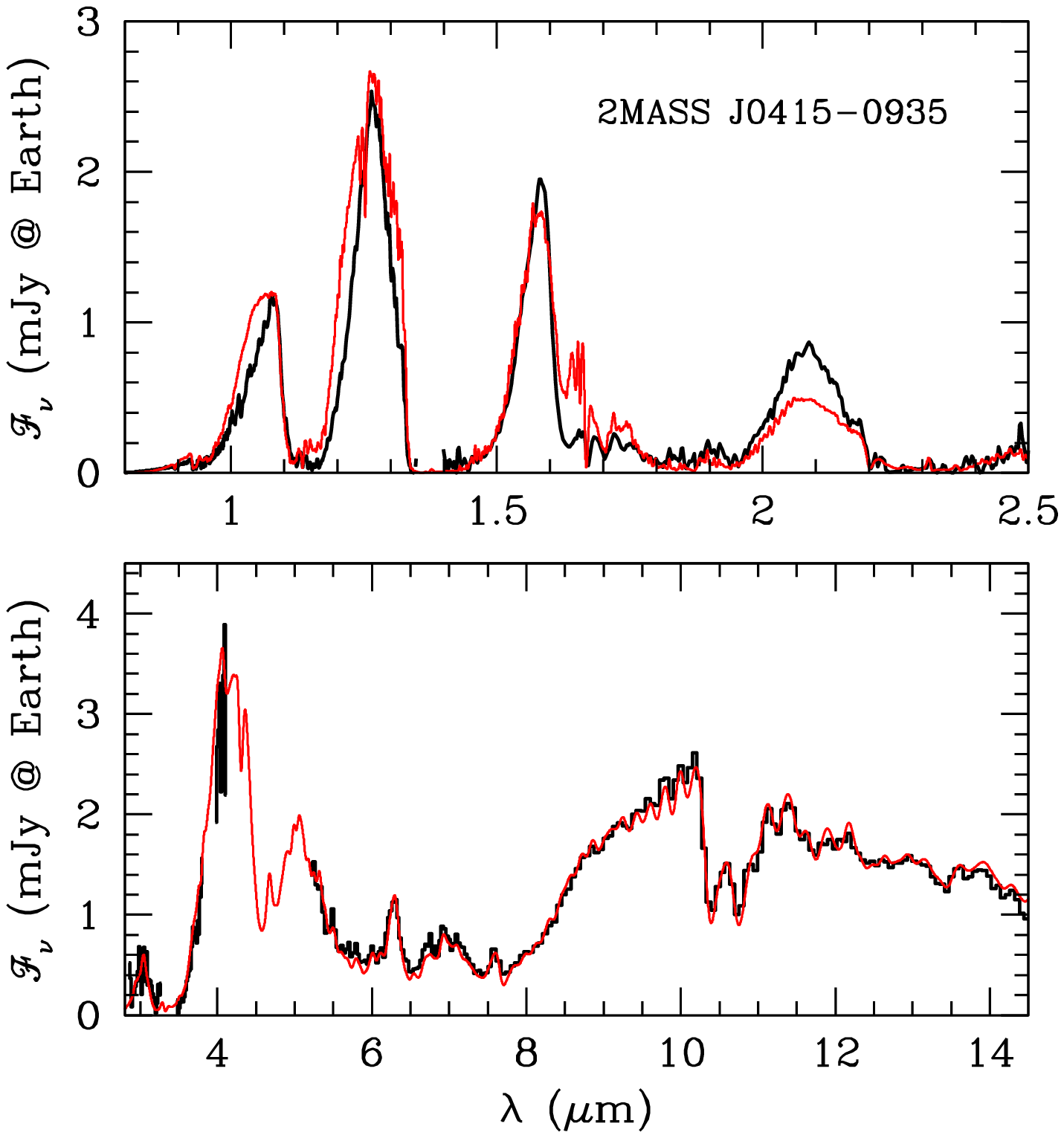}
\caption{A model spectrum that represents a typical best fit of all the spectroscopic and
photometric data for 2MASS~J0415--0935.  The model parameters are [Fe/H]=0,
$\teff=730\,$K, $\log g=5.035$
(model B in Table 2) and an eddy diffusion coefficient of $K_{zz}=10^6\,$cm$^2\,$s$^{-1}$.
Model spectra are shown at a resolving power of $\lambda/\Delta\lambda=500$ (upper panel) and 90 (lower panel) to
approximately match that of the data.  The model fluxes have not been normalized to the data
(heavy black line).
[{\it See the electronic edition of the Journal for a color version of this figure.}]}
\end{figure}

\begin{figure} \includegraphics[angle=0,scale=.80]{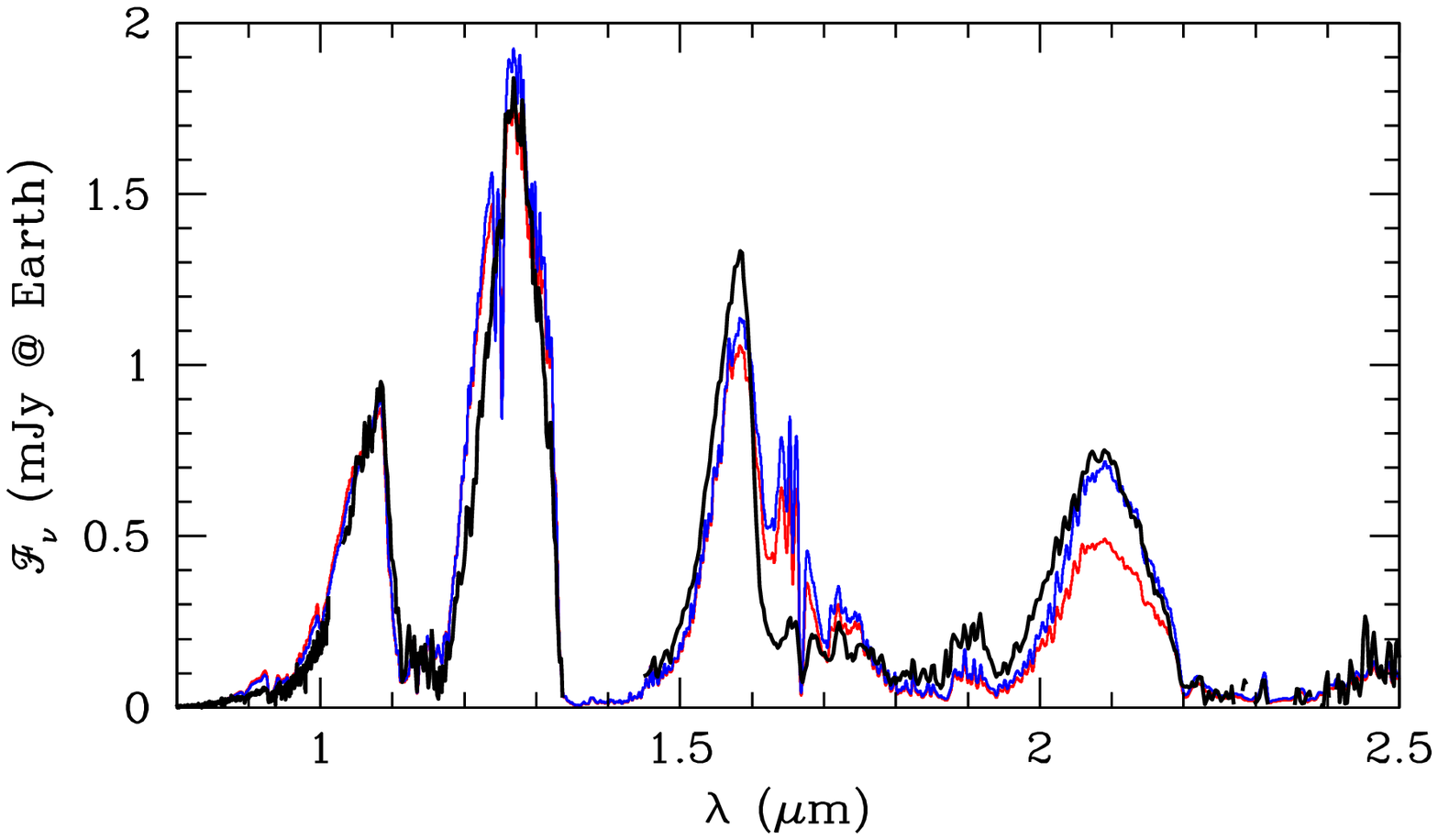}
\caption{Comparison of the optical through near-infrared spectrum of 2MASS~J1217--0311 with models
of different metallicities.  The red curve (lower thin line) is the
best fitting model with [Fe/H]=0 (model D) and the blue curve (upper thin line) is the best fitting
model with [Fe/H]=$+0.3$ (model E$^\prime$).  Both models assume chemical equilibrium ($K_{zz}=0$).
The model fluxes have not been normalized to the data but have been smoothed to a resolving power of
$\lambda/\Delta\lambda=500$ to approximate that of the data.  The data is shown by the heavy black line.
[{\it See the electronic edition of the Journal for a color version of this figure.}]}
\end{figure}

\begin{figure} \includegraphics[angle=0,scale=.80]{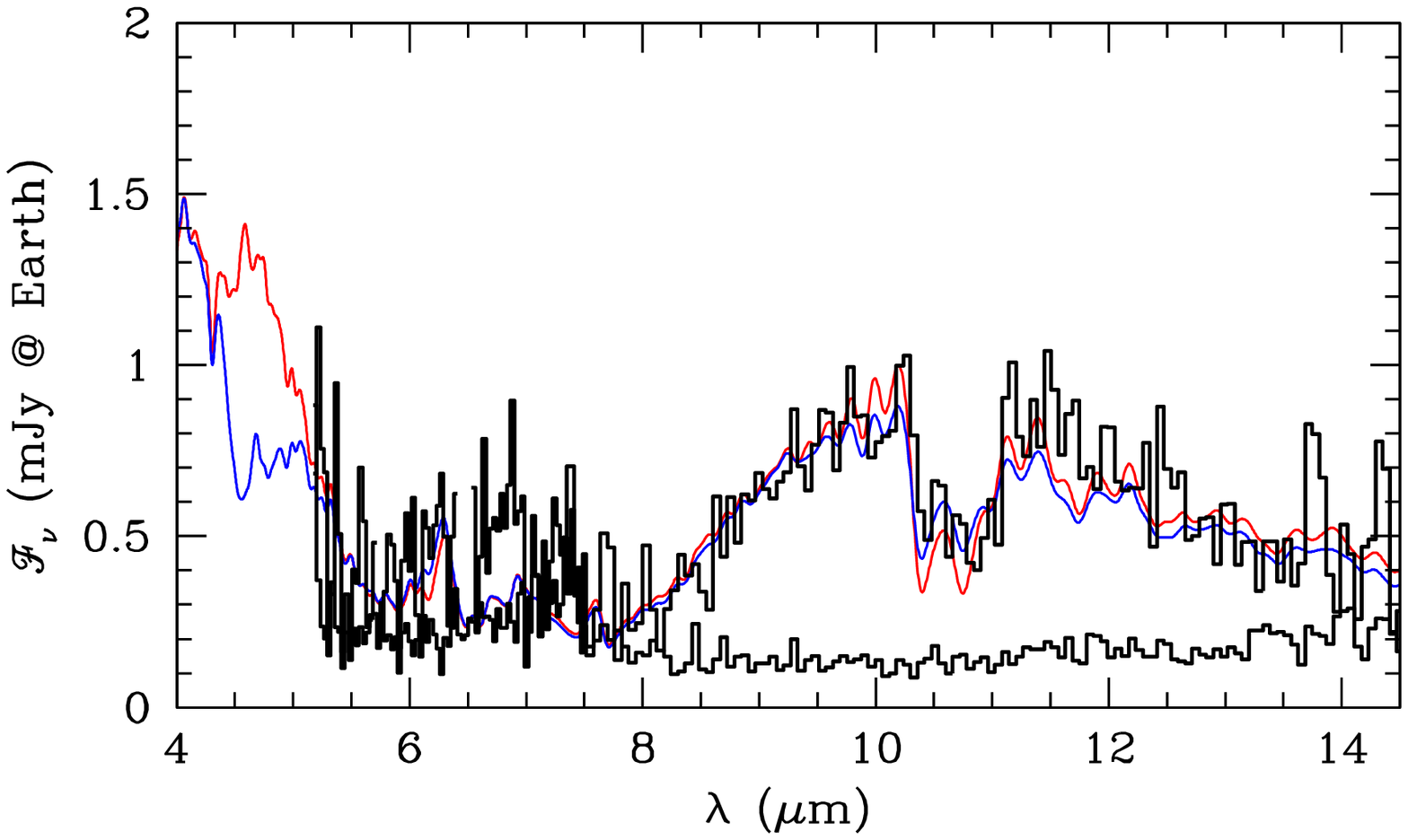}
\caption{Fits of the IRS spectrum of 2MASS~J1217--0311 showing the difference between
a model in chemical equilibrium and a model that includes vertical transport that drives
the nitrogen and carbon chemistry out of equilibrium.  The red curve (upper thin line) is the
best fitting model in chemical equilibrium (model E$^\prime$ from Table 3, with 
$K_{zz}=0$) and the blue curve (lower thin line) is a representative non-equilibrium model
(model F$^\prime$ with $K_{zz}=10^2\,$cm$^2\,$s$^{-1}$).  The data and the noise spectrum
are shown by the histograms (black).  The uncertainty on the flux calibration of the
IRS spectrum is $\pm 18$\%.
The model fluxes, which have not been normalized to the data, are shown at the
resolving power of the IRS spectrum ($\lambda/\Delta\lambda=90$).
[{\it See the electronic edition of the Journal for a color version of this figure.}]}
\end{figure}

\begin{figure} \includegraphics[angle=0,scale=.80]{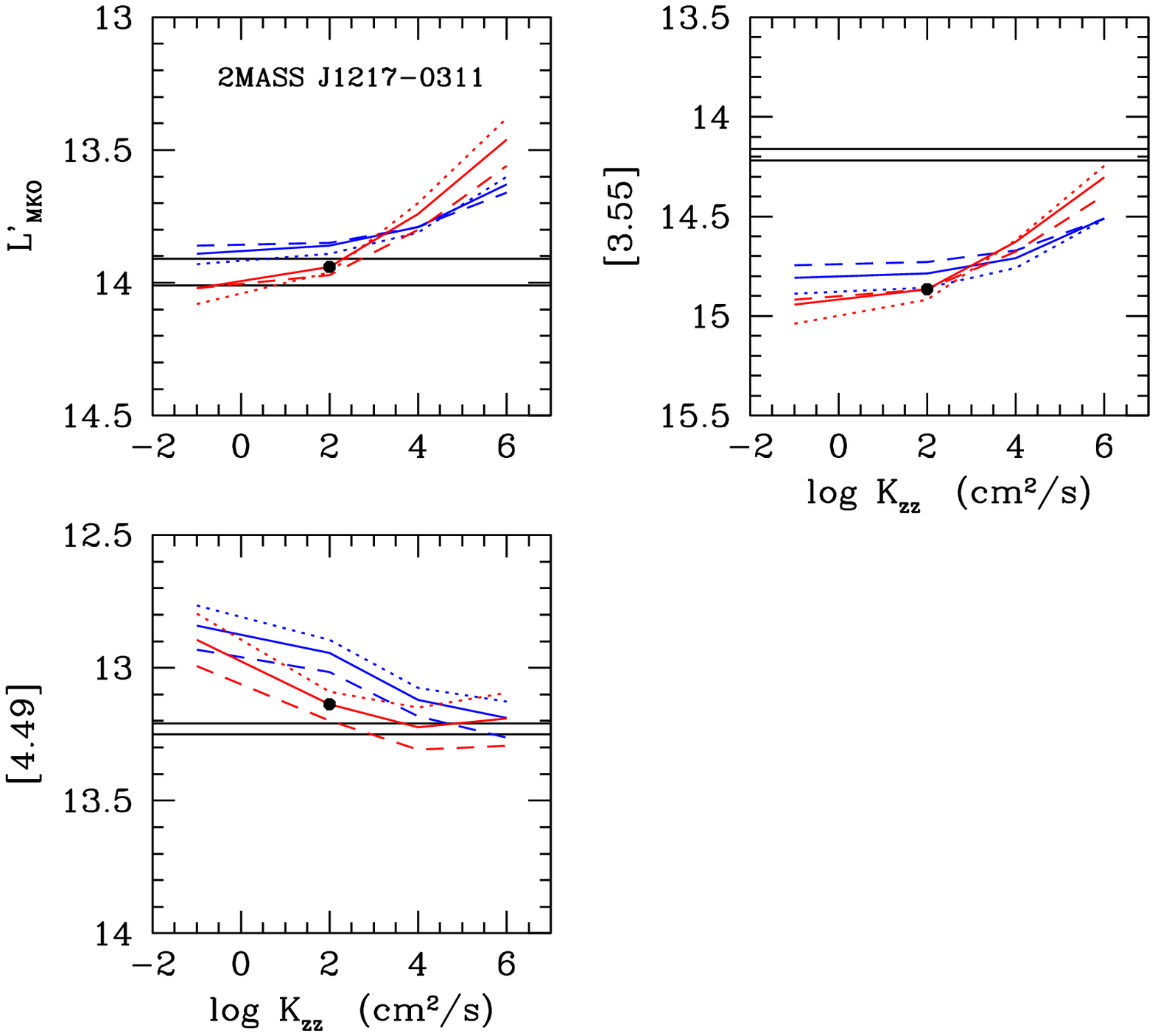}
\caption{MKO and IRAC photometry of 2MASS~J1217--0311 as a function of the coefficient of eddy diffusion
for a range of models.  Observed values ($\pm1\sigma$)
are shown by the horizontal black lines.  Synthetic photometry for each of the models in Table
3 is shown by the various curves: dotted lines: models D and D$^\prime$; solid lines: models
E and E$^\prime$; dashed lines: models F and F$^\prime$.  Blue curves are for [Fe/H]=0
and red curves for [Fe/H]=$+0.3$, the latter are below the former for small $K_{zz}$.  The solid dots
show the values for a representative best fit model (see Figure 11).  Models with $K_{zz}=0$
(in chemical equilibrium) are plotted at $\log K_{zz}=-1$.
[{\it See the electronic edition of the Journal for a color version of this figure.}]}
\end{figure}

\begin{figure} \includegraphics[angle=0,scale=.80]{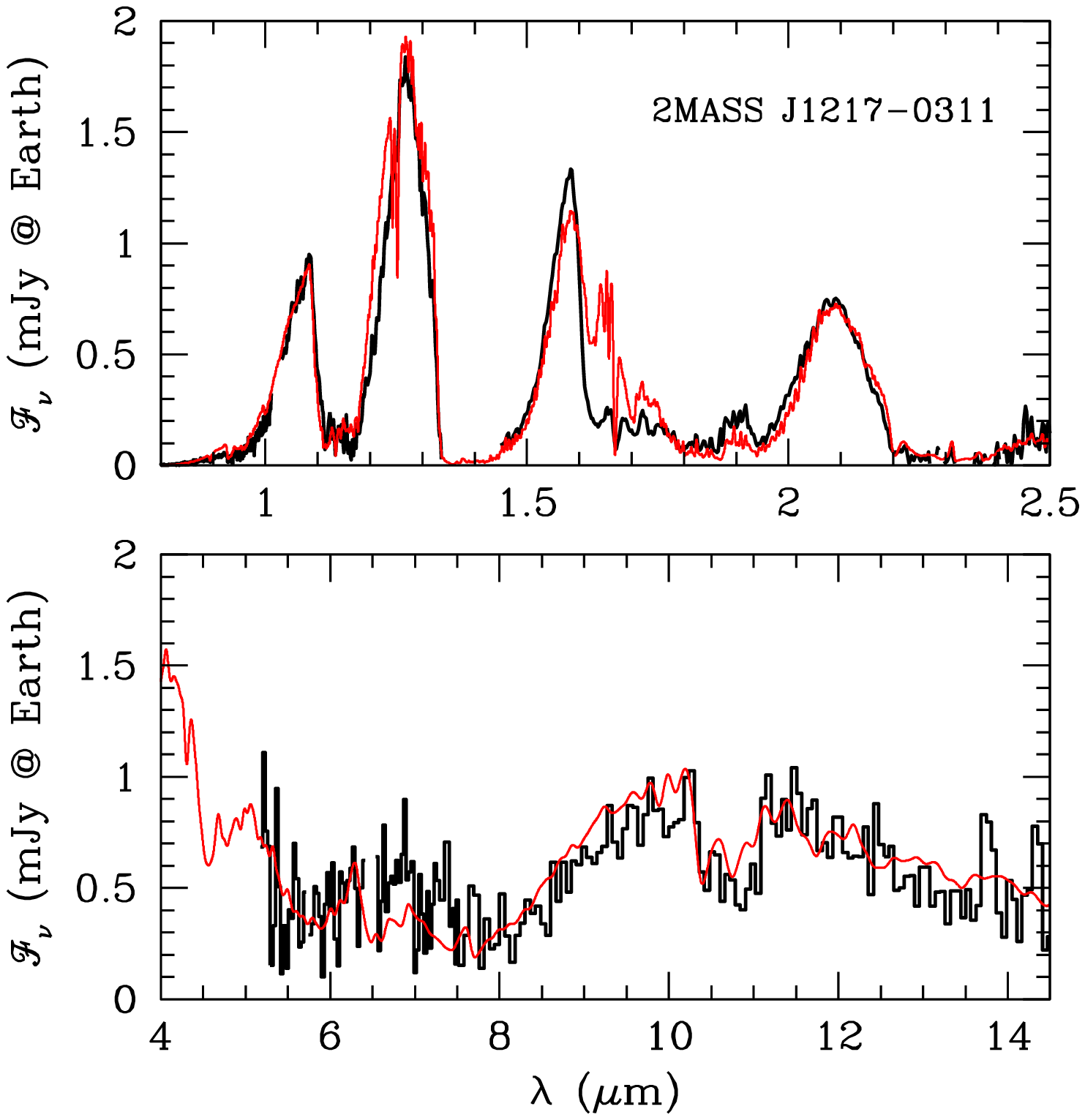}
\caption{A model spectrum that represents a typical best fit of all the spectroscopic and
photometric data for 2MASS~J1217--0311.  The model parameters are [Fe/H]=$+0.3$,
$\teff=910\,$K, $\log g=5.167$
(model E$^\prime$ in Table 3) and an eddy diffusion coefficient of $K_{zz}=10^2\,$cm$^2\,$s$^{-1}$.
Model spectra are shown at a resolving power of ($\lambda/\Delta\lambda=500$ (upper panel) and 90 (lower panel) to
approximately match that of the data.  The model fluxes have not been normalized to the data 
(heavy black curve).
[{\it See the electronic edition of the Journal for a color version of this figure.}]}
\end{figure}

\end{document}